\documentclass{article}

\usepackage{arxiv}
\usepackage{setspace, array, color, enumitem}
\usepackage{amsmath, amssymb, natbib} 
\usepackage{graphicx, booktabs}
\usepackage[english]{babel}		
\usepackage{hyperref, placeins, rotating, parskip}
\usepackage{float}
\usepackage{natbib}

\newcommand{\e}[1]{\eqref{#1}}
\newcommand{\bs}[1]{\boldsymbol{#1}}

\newtheorem{theorem}{Theorem}

\begin{document}

\title{Differential Item Functioning via Robust Scaling}
\date{July 8, 2022; Last updated November 8, 2023}

\author{Peter F. Halpin\thanks{
		Correspondence: 100 E Cameron Ave, Office 1070G, Chapel Hill, NC 27514. Ph: 919-962-7528 
		\newline \hspace{5pt}
		Acknowledgment: The author would like to thank Dr. Mattias von Davier for comments that improved the proof of Theorem 1.}  \\
	   School of Education\\
	   University of North Carolina at Chapel Hill \\
	   \texttt{peter.halpin@unc.edu}}

\hypersetup{
pdftitle={Differential Item Functioning via Robust Scaling},
pdfsubject={stat.ME, stat.AP},
pdfauthor={Peter F.~Halpin},
pdfkeywords={Item response theory, Differential item functioning, Test scaling, Robust statistics},
}

\maketitle
\begin{abstract}
This paper proposes a method for assessing differential item functioning (DIF) in item response theory (IRT) models. The method does not require pre-specification of anchor items, which is its main virtue. It is developed in two main steps, first by showing how DIF can be re-formulated as a problem of outlier detection in IRT-based scaling, then tackling the latter using methods from robust statistics. The proposal is a redescending M-estimator of IRT scaling parameters that is tuned to flag items with DIF at the desired asymptotic Type I Error rate. Theoretical results describe the efficiency of the estimator in the absence of DIF and its robustness in the presence of DIF. Simulation studies show that the proposed method compares favorably to currently available approaches for DIF detection, and a real data example illustrates its application in a research context where pre-specification of anchor items is infeasible. The focus of the paper is the two-parameter logistic model in two independent groups, with extensions to other settings considered in the conclusion.

\end{abstract}

\newpage

%  --------------------------------------------------------------------------------------------------------------
The topic of this paper is differential item functioning (DIF) in item response theory (IRT) models. The motivating application is the measurement of human development in cross-cultural contexts, which often involves translation and adaptation of existing assessments, or the development of new assessments, for use in new populations. In this context, the usual assumptions made in DIF analysis are not viable. For example, it cannot be assumed that DIF is limited to only a small proportion of items on an assessment or that a subset of items without DIF (``anchors'') can be reliably identified ahead of time. Consequently, this paper seeks to develop a method for DIF analysis that can be used in the absence of these assumptions. In particular, the paper has two main goals. 

The first goal is to consider how identification constraints on the distribution of the IRT latent trait, referred to colloquially as ``scaling'', are related to procedures used to assess DIF. This amounts to (yet another) discussion of the circular nature of DIF \citep{Angoff1982}, with the overall argument being that IRT-based scaling and DIF are two sides of the same problem. In particular, DIF with respect to a grouping variable is formally similar to IRT-based scaling with the common items non-equivalent groups (CINEG) design \citep[][chap. 6]{Kolen2014}. Items with DIF translate into outliers in the CINEG design. The latter problem has received some attention in the scaling literature \citep{He2015, He2020, Stocking1983}, although the potential advantages of this approach for DIF analysis seem to have gone largely unnoticed. The second goal of this paper is to elaborate on these advantages.  

Framing DIF in terms of outlier detection allows for the theory of robust statistics to be brought to bear on the problem. The general strategy taken in this paper is to approach IRT-based scaling with the CINEG design from the perspective of M-estimation of a location parameter \cite[][chap. 4]{Huber2009}. In this context, the item parameter estimates play the role of data points whose location we wish to estimate. Whereas standard M-estimation theory involves asymptotics in the number of data points (i.e., items), the results developed in this paper invoke asymptotics in the number of respondents in the IRT model while treating the number of items as fixed. Taking this approach, the asymptotic distribution of a relatively wide class of M-estimators of IRT-based scaling parameters is obtained. These results are then used to construct a highly robust redescending M-estimator that can be tuned to flag items with DIF at the desired asymptotic Type I Error (false positive) rate. 

It is shown that the finite sample breakdown point (FSBP) of the proposed estimator depends on only the choice of starting value. This is unlike more typical M-estimation problems in which one must also consider the breakdown of an ancillary estimate of the scale (variance) of data points \cite[][chap. 6]{Huber2009}. As a consequence, the proposed estimator can be constructed to achieve the theoretical maximum FSBP for any translation equivariant estimator, which is 1/2 \cite[][\S 11.2]{Huber2009}. This means that the proposed estimator remains bounded whenever fewer than 1/2 of the items on a test exhibit DIF. Theoretical guarantees about FSBP are quite weak, so these results are complemented by data simulations illustrating how the proposed procedure, as well as some currently available DIF detection methods, perform ``on the way to breakdown." A second simulation focuses on statistical power, and the simulations are followed by a real data example from cross-cultural human development in which the pre-specification of anchor items is infeasible. 

The proposed methodology is referred to as the Robust DIF (R-DIF) procedure. Its main advantages are that (a) anchor items need not be identified ahead of time, and (b) theoretical and simulation-based results provide some assurances about its performance when fewer than 1/2 of the items on an assessment exhibit DIF. It can be implemented as a post-estimation procedure following separate calibrations of a focal IRT model in the populations of interest, and it does not require multi-group models or iteratively fitting models with different parameter restrictions. Standard computational procedures can be used for estimation (e.g., iteratively reweighted least squares), and have been implemented in an R \citep{R} package accompanying this paper, \texttt{robustDIF}, which is briefly described. 

The focus of the paper is the two-parameter logistic (2PL) model in two independent groups. The main developments address DIF in the item difficulty parameter, which simplifies presentation. Extensions to the item discrimination parameter are then shown to follow directly from the main results. The next section reviews the literature with the purpose of making connections between DIF, IRT-based scaling, and robust statistics. While many of these issues extend beyond the context of IRT, the focus of the review is IRT-based methods. 

%-----------------------------------------------------------------------------------------
\section{The Circular Nature of DIF: Redux} 

DIF involves two interrelated problems. The first and more obvious problem is to infer whether item parameters differ as a function of one or more external variables. One way to do this is Lord's (\citeyear{Lord1980}) test, which is considered here for illustrative purposes. Let $p_{gi}$ denote the probability of a respondent in group $g = 0, 1$ endorsing a binary item $i = 1, \dots, m$. Specify the 2PL IRT model as:
\begin{equation} \label{IRT1}
 \text{logit}(p_{gi})  = a_{gi} \, (\eta_{g} - b_{gi}) \; \text{ with } \; \eta_g \sim N(\mu_g, \sigma^2_g),
\end{equation}
where $a_{gi}$ is the item discrimination parameter, $b_{gi}$ is the item difficulty parameter, and $\eta_{g}$ is the latent trait. Then Lord's test for the item difficulty parameters can be written as
\begin{equation} \label{lord} \notag
z_i = \frac{\hat b_{1i} - \hat b_{0i}} {\sqrt{\text{var}(\hat b_{1i}) + \text{var}(\hat b_{0i})}}.
\end{equation}
When $\hat b_{gi}$ is the maximum likelihood estimate (MLE) of $b_{gi}$, $z_i$ is a Wald test. 

The second problem has to do with the identification of IRT models. In particular, the 2PL IRT model is identified only up to an affine transformation of the latent trait \citep[see][ \S 2.2.3] {vanderLinden2016}. Thus, using the transformed values $\eta_g^* = A \eta_g + B$, $b_{gi}^* = A b_{gi} + B$, and $a_{gi}^* = a_{gi} / A$ in Equation \e{IRT1} leaves $\text{logit}(p_{gi})$ unchanged. A common way to address this problem is by setting $\mu_g$ and $\sigma^2_g$ to fixed values, which is referred to as scaling the latent trait.  

Each of these two problems has implications for the other.  Let us first consider the implications of the identification problem for testing the item parameters. In the DIF literature, differences in the distribution of the latent trait across groups are referred to as ``impact'' \citep{Angoff1993}. When the latent trait is scaled separately in each group, this requires that we ignore impact. In particular, assuming that the latent trait has the same distribution in both groups is equivalent to transforming the latent variable as
\begin{equation} \label{tstar} \notag
 \eta^*_1 = \sigma_0 \left(\frac{\eta_1 - \mu_1}{\sigma_1}\right) + \mu_0. 
 \end{equation}
The corresponding transformation of the item difficulty parameters is
\begin{equation} \label{dstar} 
b^*_{1i} = \sigma_0 \left(\frac{b_{1i} - \mu_1}{\sigma_1}\right) + \mu_0,
\end{equation} 
and plugging $b^*_{1i}$ into Lord's test gives
\begin{align} \label{zstar}
 z^*_i & = \frac{\hat b^*_{1i} - \hat b_{0i}} {\sqrt{\text{Var}(\hat b^*_{1i}) + \text{Var}(\hat b_{0i})}}  
             = \frac{\frac{\sigma_0}{\sigma_1} \left(\hat b_{1i} - \mu_{1}\right) + \mu_0 - \hat b_{0i}} {\sqrt{\frac{\sigma^2_0}{\sigma^2_1} \text{Var}(\hat b_{1i}) + \text{Var}(\hat b_{0i})}}. 
\end{align}
Equation \e{zstar} shows how the identification problem affects Lord's test: When impact is ignored, Lord's test is biased by the mean and variance of the latent trait in both groups. Stated more generally, comparing item parameters over groups requires solving the scaling problem. 

In the context of two independent groups, the usual way to solve the scaling problem is to (a) arbitrarily scale the latent trait in only one of the groups, and then (b) assume that some (or all) of the item parameters are equal over groups. Part (a) is warranted because, as noted, the 2PL model is identified only up to an affine transformation of the latent trait. Part (b) then suffices to scale the latent trait in the second group, for example, by setting $b_{1i} = b_{0i}$ for at least two items and then solving for $\mu_1$ and $\sigma_1$ using Equation \e{dstar}. In the literature on IRT-based scaling, this two-part approach is referred to as the CINEG design \cite[see][\S 6.3.1]{Kolen2014}. %The usual application arises when each group of respondents receives a different assessment, but the assessments have some items in common. %In the CINEG design, the common items are assumed not to exhibit DIF with respect to the grouping variable. 
In the literature on DIF, part (b) of the scaling problem is referred to as choosing an ``anchor set'' of items \cite[][]{Kopf2015b}. Choosing anchor items brings us back to the problem of comparing item parameters over groups, whence the circular nature of DIF \citep{Angoff1982}.    

Traditional approaches to DIF analysis sought to circumvent this issue by proceeding iteratively, first assuming an anchor set, then testing DIF on each item, then updating the anchor set, and so on. The two-stage ``purification'' and ``refinement'' approach \citep{Dorans1993} is perhaps the best-known example of this strategy. While this two-stage approach can work well in some settings, it does not control Type I Error rates when a moderate proportion of items (e.g., $\geq 1/4$)  are biased in the same direction \cite[e.g.][]{Kopf2015a}. While many alternative strategies for selecting anchors have been proposed, these rely mainly on heuristic arguments about the size of the anchor set and criteria for selecting anchors \cite[e.g.][]{Kopf2015b, Kopf2015a}. In general, approaches based on anchor item selection are unsatisfying from a theoretical perspective because subsequent tests of DIF proceed as if the anchors were known a' priori. 

Becgher and Maris (\citeyear{Bechger2015}) proposed a test of DIF that is invariant under affine transformation of the latent trait. This approach avoids the logical circularity of traditional methods, but results in pairwise comparisons over items, rather than a test of individual items, which is a practical shortcoming. Yuan and colleagues (\citeyear{Yuan2021}) proposed a Monte-Carlo test that replaces pairwise comparisons over items with comparison to a single reference point, although the latter is taken as the average of anchor items whose selection remains largely heuristic. Other recent research has used regularization methods to simultaneously estimate item parameters and scaling parameters, while imposing sparsity on quantities that govern DIF \citep{Belzak2020, Magis2015, Schauberger2020}. Although the use of regularization for variable selection in regression-based models is well established, the theoretical motivation for using regularization to address the scaling of IRT models is less clear. The simulation studies presented in this paper suggest that the performance of regularization-based approaches is not qualitatively different from traditional DIF methods. 

In what follows I propose an alternative approach to DIF analysis. The overall idea is to tackle the scaling problem directly using robust statistics. Early work on IRT-based scaling considered, and dismissed, approaches based on robust statistics \cite[][Appendix]{Stocking1983}. He and colleagues \citep{He2013, He2015, He2020} revisited the topic of robust scaling, and this work is a source of inspiration for the present research. In particular, He (\citeyear{He2013}) considered outlier detection and omission in the context of IRT-based scaling, but dismissed this approach with the rationale of preserving content coverage in the anchor set. Recently, an independent line of research by Wang and colleagues (\citeyear{Wang2022}) has addressed the use of robust regression in DIF analysis. This present paper focuses on the related problem of IRT-based scaling via M-estimation of a location parameter, contributing a relatively general asymptotic test of DIF as well as theoretical results on the robustness of the proposed R-DIF procedure. Other recent work has applied robust methods to the choice of cut-off values for item fit indices \citep{vonDavier2022}, but has been developed under the assumption that only a small proportion of items may exhibit DIF. 

Recent research has also emphasized the connection between scaling and DIF \citep{Doebler2019, Stenhaug2021, Strobl2021}, although these approaches have not made use of robust methods to address the scaling problem. Another related approach is the alignment procedure \citep{Asparouhov2014, Robitzsch2023}, in which the configural model is estimated as a first step, and then a loss function is used to minimize the degree to which item parameters vary over groups. R-DIF is also a post-estimation procedure, but the goal is not to minimize DIF – rather, the goal is to obtain a robust estimate of IRT scaling parameters and use this as a basis for testing for DIF. 
 
Although the theory of M-estimation is well established \cite[][]{Huber2009}, there are some peculiar aspects of the IRT-based scaling problem that do not feature in the more general theory and therefore warrant special attention. First, the population model in IRT-based scaling is known a' priori (e.g., Equation \e{dstar}). Knowing the population model means that we can aggressively pursue model-based outlier detection, without worrying about whether the model is correct. Second, the population model is deterministic (e.g., unlike regression models, the linear relationship in Equation \e{dstar} does not contain a residual term). This implies that the only source of variation in the sample-based scaling problem is the (co-)variances of the item parameters estimates, which are available, for example, via known results on maximum likelihood estimation in IRT \cite[e.g.,][]{Bock2021}. Third, asymptotic results for the proposed M-estimator can be obtained via the IRT parameter estimates, and this provides an alternative route to inference than is usually considered in M-estimation theory. These details are elaborated in the following section.

%Taken together, these peculiar details lead to a novel approach for provide an extraordinarily strong starting point from which to build up a methodology for DIF analysis. This intuition is elaborated in the following sections. 

%-----------------------------------------------------------------------------------------
\section{The R-DIF Procedure} 

The overall logic of the R-DIF procedure is to obtain a robust estimator of IRT scaling parameters and then use it to construct a robust test of DIF. It turns out that tuning the estimator to be robust to DIF is tantamount to flagging (down-weighting) items with DIF during estimation. Additionally, both the estimator and test can be parameterized such that the only quantity directly affected by DIF is the IRT scaling parameter itself. This is importantly different from similar problems in M-estimation that require an ancillary estimate of the scale (variance) of the data, and is the key to the robustness (i.e., high breakdown point) of the R-DIF procedure. The main steps involved in developing the R-DIF procedure are summarized below, and the remainder of this section describes each step in more detail.  

The R-DIF procedure is based on an M-estimator of a location parameter, $\theta$. The estimator, $\tilde \theta$, can be defined in terms of the estimating equation
\begin{equation} \label{M} 
\Psi(\theta) = \sum_{i=1}^{m} \psi \left(\frac{Y_i - \theta}{s_i} \right) = 0, 
\end{equation} 
which is developed in the following three steps: 
 
\begin{itemize}

\item[1.] $Y_i$ is defined as a scalar-valued function of the MLEs of the parameters of item $i = 1, \dots, m$, such that $\sqrt{n}\, (Y_i  - \theta) \overset{d}{\rightarrow} N (0, \tau_i)$ under the null hypothesis that item $i$ does not exhibit DIF, with $n$ denoting the number of respondents. The $Y_i$ play the role of the sample data points whose location parameter we wish to estimate.  

\item[2.] For a relatively general choice of $\psi$, choosing $s_i = \tau_i$ to be the variance of the asymptotic null distribution of $Y_i$ is shown to lead to an efficient estimator of $\theta$ as well as a convenient asymptotic test of DIF. 

\item[3.] The loss function $\psi$ is chosen to be a so-called \emph{redescending}  function \cite[][\S. 4.8]{Huber2009} that is tuned so that values of $|Y_i - \tilde\theta| / \tau_i$ beyond the $1-\alpha/2$ quantile of its asymptotic null distribution are automatically set to zero during estimation. 

\end{itemize}

%-----------------------------------------------------------------------------------------
\subsection{Step 1: Setting up the scaling problem} 

Specify the IRT models as above, although now in slope-intercept form, in the reference and comparison groups, respectively:
\begin{align} \notag \label{IRT}
 \text{logit}(p_{0i}) & = a_{0i} \eta_{0} + d_{0i} \; \text{ with } \; \eta_0 \sim N(0, 1) \\ 
\text{logit}(p_{1i}) & = a_{1i} \eta_{1} + d_{1i} \; \text{ with } \; \eta_1 = (\eta^*_1  - \mu) / \sigma \; \text{ and } \;  \eta^*_1 \sim N(\mu, \sigma^2). 
\end{align}
The scaling problem requires solving for $\mu$ and $\sigma$ using the relation \cite[cf.][\S 6.2.1]{Kolen2014} 
\begin{align}  \notag
 a_{1i} \eta_{1} + d_{1i} & = a^*_{1i} \eta^*_{1} + d^*_{1i},
\end{align}
from which follows the scaling equations:
\begin{align} \label{CINEG1}
\sigma & = a_{1i} / a^*_{1i} \\ \label{CINEG2}
\mu / \sigma & = (d_{1i} - d^*_{1i}) / a_{1i}.
\end{align}
In the CINEG design, we let the item parameters in the reference group stand in for the ``un-scaled'' item parameters in the comparison group:
\begin{align}  \label{CINEG3}
 a^*_{1i} & = a_{0i} \\ \label{CINEG4}
 d^*_{1i} & = d_{0i}.  
\end{align}
These two equalities assert that item $i$ does not exhibit DIF with respect to group membership, which makes explicit the formal connection between IRT-based scaling using the CINEG design and DIF. I will refer to Equations \e{CINEG3} and \e{CINEG4} as null hypotheses about DIF on the slope and intercept of item $i$, respectively. 

Although it is usual to isolate $\mu$ by substituting for $\sigma$ in Equation \e{CINEG2}, it is preferable to treat $\theta = \mu/ \sigma$ as the target parameter when addressing DIF. Taking this approach, it can be seen that the null hypothesis about item slopes in Equation \e{CINEG3} applies only to the scaling equation for $\sigma$ in Equation \e{CINEG1}. Similarly, the null hypothesis about item intercepts in Equation \e{CINEG4} applies only to the scaling equation for $\theta$ in Equation \e{CINEG2}. This situation can be contrasted with the more usual distinction between uniform and non-uniform DIF \citep{Mellenbergh1982}. In particular, uniform DIF evaluates the item difficulties (rescaled intercepts) under the assumption that there is no DIF on the item slopes, whereas Equation \e{CINEG2} does not require this assumption. This is convenient because it allows for DIF in each type of item parameter to be addressed separately. In what follows I focus on the item intercepts only (i.e., Equations \e{CINEG2} and \e{CINEG4}). The same developments apply to the item slopes with only minor modifications, and it will also be shown how to test both item parameters simultaneously; these topics are deferred to the section of this paper entitled ``Extensions''. 

In practice, the item parameters will be estimated in independent samples of sizes $n_0$ and $n_1$. To set up the sample-based problem, collect the parameters of item $i$ in the vector $\bs{\nu}_i = [a_{0i}, d_{0i}, a_{1i}, d_{1i}]^T$, let $\bs{\nu} = [\bs{\nu}_1^T, \bs{\nu}_2^T, \dots, \bs{\nu}_m^T]^T$, and write 
\begin{equation} \label{Y}
Y_i(\bs{\nu}) =  (d_{1i} - d_{0i}) / a_{1i}. 
\end{equation}
In what follows, it is assumed that MLEs $\hat{\bs{\nu}}$ and their asymptotic covariance matrix $\text{cov}({\hat {\bs{\nu}}})$ are available \cite[see e.g.,][]{Bock2021}. The shorthand notation $Y_i = Y_i(\hat{\bs{\nu}})$ is reserved for the sample-based quantities only. 

For later reference, it is noted that application of the Delta method yields \cite[see e.g.,][]{Vaart1998}:
\begin{equation} \label{delta1}
\sqrt n (Y_i - Y_i(\bs{\nu}) ) \overset{d}{\rightarrow} N(0, \text{var}(Y_i ))
\end{equation}
where $n = n_0 + n_1$, $n_0 / n_1 = c$ for $ c \in (0, \infty) $, and  
\begin{equation} \label{delta2} 
\text{var}(Y_i ) = \nabla Y_i(\bs{\nu})^T \text{cov}({\hat {\bs{\nu}}}) \nabla Y_i(\bs{\nu}). 
\end{equation}
%
%Throughout this paper the notation $\text{var}(U)$ will be used only to denote the asymptotic variance of $U$ as $n \rightarrow \infty$. 
%
Under the null hypothesis in Equation \e{CINEG4}, we have $Y_i(\bs{\nu}) = \theta$ (via Equation \e{CINEG2}) so that $E(Y_i) = \theta$ and the gradient elements corresponding to item $i$ are:
\begin{align} \label{grad} \notag
 \nabla Y_i(\bs{\nu}_i) & = a_{1i}^{-1} \left[0, -1 , - {Y_i(\bs{\nu})}, \; 1\right]^T\\
			       & = a_{1i}^{-1}  \left[0, -1,  - \theta, \; 1 \right]^T. 
\end{align}
Note that the other gradient elements are equal to zero. Thus the ``null variance'' of $Y_i$ can be written as a function of $\theta$, say $\tau_i = \tau_i(\theta)$, with Equations \e{delta2} and \e{grad} leading to the explicit expression
\begin{equation} \label{tau} 
\tau_i(\theta) \equiv \text{var}(Y_i ) = a_{1i}^{-2} \left(\theta^2 \text{var}(\hat a_{1i}) - 2 \theta \,\text{cov}(\hat a_{1i}, \hat d_{1i}) + \text{var}(\hat d_{1i}) + \text{var}(\hat d_{0i}) \right).
 \end{equation}

The foregoing results provide a key idea behind this paper: The asymptotic null distribution of $Y_i$ can be obtained by using $\theta$ in place of $Y_i$. Indeed, when comparing the R-DIF procedure to previous work on robust scaling \cite[e.g.,][]{Stocking1983, He2013, Wang2022}, the substitution in the second line of Equation \e{grad} is perhaps the crucial difference. To anticipate Theorem 2 below, treating $\tau_i$ as a function of $\theta$ allows for the R-DIF procedure to achieve a FSBP of 1/2. In practice, this means that we can obtain a reasonable estimate of the null distribution of $Y_i$, so long as fewer than one-half of the items exhibit DIF. 

%-----------------------------------------------------------------------------------------
\subsection{Step 2: Choosing the weights $s_i$.}

One rationale for choosing the weights $s_i$ in Equation \e{M} is to ensure that the resulting M-estimator has acceptable efficiency in the absence of outliers \cite[][\S 2.3.2]{Maronna2019}. In the present context, the absence of outliers corresponds to the joint null hypothesis that none of the items exhibit DIF.  The first part of Theorem 1 below shows that, for a relatively general choice of $\psi$, setting $s_i = \tau_i$ results in an unbiased and asymptotically (in $n$) efficient estimator of $\tilde \theta$, under the joint null hypothesis that none of the item intercepts exhibit DIF. The second part of the theorem obtains the distribution of $Y_i - \tilde\theta$ under these same conditions. Subsequent remarks address the utility of these results in the context of DIF analysis. 

The following notation is required. Let the function $\theta = \theta(\bs{\nu})$ be implicitly defined by 
\begin{equation} \label{M2} 
\Psi(\bs{\nu}, \theta) = \sum_{i=1}^{m} \psi \left(U_i(\bs{\nu})\right) = 0
\end{equation} 
with $U_i(\bs{\nu}) = (Y_i(\bs{\nu}) - \theta)/s_i$ and $s_i > 0$. The M-estimator computed using the IRT MLEs $\hat{\bs{\nu}}$ will be wrtitten $\tilde \theta = \theta(\hat{\bs{\nu}})$. The following assumptions about $\psi$ are also required: 
\begin{itemize} 
\item[\hspace{10pt} A1] $\psi$ is differentiable with $\psi' = d\psi(u) / du$.
\item[A2] For some positive constant $k$ and $u \in (-k , k)$, $\psi(u) = 0$ if and only if $u = 0$. 
\item[A3] $\psi'(0) = c > 0$. 
\item[A4] $ \Psi' = \partial \Psi / \partial \theta \neq 0$ at $\theta_0 = \mu /\sigma$. 
\end{itemize}

Assumptions (A1) through (A3) are not restrictive for many common choices of $\psi$ \cite[e.g., see][chap. 2]{Maronna2019}, although they do exclude some more robust choices, notably the median (by A1). Assumption (A4) is required to obtain the general (i.e., non-null) asymptotic distribution of $\tilde\theta$ and of $Y_i - \tilde\theta$ for non-monotone $\psi$, but is not restrictive for the null distribution (see Equation \e{w2} in the Appendix). M-estimators of location are typically characterized by the following additional assumption, which is noted here for later reference but is not required by the theorem:
\begin{itemize} 
\item[A5] $\psi(u)$ is odd. 
\end{itemize} 
%
%------------------------------------------------------------------------
\begin{theorem}
For the two-group IRT model in Equation \e{IRT}, let $\theta_0 = \mu/\sigma$ denote the target scaling parameter, with item parameters collected in the vector $\bs{\nu}$, and MLEs $\sqrt{n}\, (\hat{\bs{\nu}} - \bs{\nu}) \overset{d}{\rightarrow} N(\bs{0}, \text{cov}(\hat{\bs{\nu}}))$ obtained in a sample of size $n = n_0 + n_1$, with $n_0/n_1 = c$ for $c \in (0, \infty)$. Let $\theta(\bs{\nu})$ and $\tilde \theta = \theta(\hat{\bs{\nu}})$ be defined as in Equation \e{M2} and Assumptions A1 - A4. Assume the joint null hypothesis that $Y_i(\bs{\nu}) = \theta_0$ for $i = 1, \dots, m$ (i.e., no DIF). Then choosing $s_i = \text{var}(Y_i)$ implies

\noindent Part \emph{(a):} $\sqrt n\, (\tilde \theta - \theta_0) \overset{d}{\rightarrow} N(0, \text{var}(\tilde \theta))$ where
\begin{equation} \notag
	\text{var}(\tilde \theta) = \frac{1}{ \sum_i \text{var}(Y_i)^{-1}}.
\end{equation}

is a lower bound on the variance of $\tilde \theta$. \\
\noindent Part \emph{(b):} $\sqrt n\,  T_i \overset{d}{\rightarrow} N(0, 1)$ with
\begin{equation} \notag
T_i =  \frac{Y_i -  \tilde \theta }{\sqrt{\text{var}(Y_i) - \text{var}(\tilde \theta)}}. 
\end{equation}

\end{theorem} 
%-----------------------------------------------------------------------------------------

The proof is given in the Appendix. Part (a) describes the asymptotic distribution of the estimated scaling parameter $\tilde \theta$, under the joint null hypothesis that no items exhibit DIF. In particular, it shows that setting $s_i = \text{var}(Y_i)$ ensures that the resulting estimator $\tilde \theta$ is asymptotically efficient in the absence of DIF. This may seem counterintuitive in light of well-known results about the relative inefficiency of robust estimators \cite[e.g.,][chap. 6]{Huber2009}. However, past results use asymptotics in $m$, which do not feature in the current approach. In the present case, the intuition is as follows. As $n \rightarrow \infty$, the null distribution of $Y_i$ becomes increasingly concentrated around $\theta_0$ for each $i = 1, \dots, m$. Thus, in the limit, the loss function $\psi$ influences the null distribution of $Y_i$ only via the constant $\psi'(0)$ (see A3), and this constant cancels out when computing the variance of $\tilde \theta$ (see Equation \e{w1}). 

Part (b) of the theorem provides a Wald test of DIF for a relatively wide class of M-estimators of $\theta$. The result shows that a robust test will require robust estimates of both $\theta$ and $\text{var}(Y_i)$, $i = 1, \dots, m$. However, as noted in connection with Equation \e{tau}, these are one and the same problem. In particular, under the null hypothesis of no DIF on item $i$, $\text{var}(Y_i) = \tau_i(\theta)$ depends only on $\theta$ and the covariance matrix of the item's parameter estimates. Thus, a robust test of DIF requires only a robust estimate of $\theta$. The following section addresses how to obtain such an estimate. 

A final remark concerns the non-null distributions of $\tilde\theta$ and $Y_i - \tilde\theta$. The Appendix shows that both the mean and variance of these distributions depend directly on $\psi$. The study of these distributions presents interesting possibilities for future research, but will not be addressed in this paper. The simulation studies presented below provide some empirical examples of the statistical power of the R-DIF procedure. 

%-----------------------------------------------------------------------------------------
\subsection{Step 3: Choosing the loss function $\psi$. }

This section introduces additional assumptions about $\psi$ in order to obtain a robust estimator of $\theta$. A general strategy is to choose $\psi$ so that the influence of any individual data point is bounded \cite[][\S 1.5]{Huber2009}:  

\begin{itemize} 
\item[A6] $\mid \psi(u) \mid  <  c$ for some positive constant $c$.  
\end{itemize}  
This strategy is taken one step further by so-called \emph{redescending} M-estimators, which, in addition to being bounded, assign outliers a weight of zero \cite[][\S 4.8]{Huber2009}: 
\begin{itemize} 
\item[A7] $\psi(u)  = 0$ for $|u| > k$. 
\end{itemize}  
The constant $k$ is a tuning parameter that ensures the estimator is resistant to outliers $|u| > k$, and, as a bi-product, also automatically ``flags'' any such outliers during estimation. 

The usual application of redescending M-estimators is to guard against gross outliers, while also ensuring acceptable efficiency in the absence of outliers. This leads to choices of $k$ that are intended to flag only a small proportion of data points \cite[e.g.,][]{Maronna2019, vonDavier2022}. In the research context described at the outset of this paper, the goal can be better characterized in terms of guarding against potentially many modest outliers, and part (a) of Theorem 1 shows that the choice of $\psi$ does not affect asymptotic (in $n$) efficiency in the absence of outliers. Thus it is proposed to pursue a more aggressive choice of $k$. 

In particular, part (b) of Theorem 1 shows how to choose item-specific tuning parameters $k_i$ such that items with DIF are flagged at a chosen asymptotic Type I Error rate, $\alpha$. Letting $u = U_i = (Y_i - \tilde \theta) / \tau_i$, the theorem implies that $\sqrt n \, U_i  \overset{d}{\rightarrow} N(0, \omega_i)$, where
\begin{equation} \label{omega}
\omega_i  = \frac{\tau_i - \bar{\tau} / m }{\tau_i^2} %\frac{\text{var}(Y_i) - \text{var}(\tilde \theta)}{\text{var}(Y_i)^2
\end{equation}
with $\tau_i$ defined in Equation \e{tau} and $\bar{\tau}$ denoting the harmonic mean over $i = 1, \dots, m$. 
By definition, choosing $k_i$ to be the $1-\alpha/2$ quantile of $N(0, \omega_i)$ implies that $\text{Prob}(|U_i| > k_i) = \alpha$ under the joint null hypothesis that no items exhibit DIF. Using  A7, the function $\psi$ is set to zero for values of $|U_i| > k_i$. Thus, the proposed choice of $k_i$ is seen to be equivalent to an asymptotic test of DIF with size $\alpha$ -- i.e., items that exhibit DIF are flagged by setting $\psi(u_i) = 0$ during estimation of $\tilde \theta$.
 
%Then choosing $k_i$ to be the $1- \alpha/2$ quantile of $N(0, \omega_i)$ implies that $\psi(U_i) = 0$ with asymptotic probability $\alpha$, under the joint null hypothesis that none of the item intercepts exhibit DIF. Thus, the proposed choice of $k_i$ is seen to be equivalent to an asymptotic test of DIF with size $\alpha$. 

Motivating a per-item tuning parameter in terms of the desired Type I Error rate for DIF detection is a primary advantage of the proposed approach compared to that of Wang and colleagues (\citeyear{Wang2022}). In particular, the simulation studies reported below show that the R-DIF procedure yields a test of DIF that maintains the nominal value of $\alpha$ quite well even when a relatively large proportion of items exhibit DIF. 

While there are various redescending M-estimators available, Tukey's bisquare is well suited to the present context. It can be defined as: 
\begin{equation} \label{psi}
\psi(u) = \left\{\begin{array}{ccc}
u \left(1 - \left( \frac{u}{k} \right)^2\right)^2 & \text{ for   } & \mid u \mid \leq k \\		
 0 & \text{ for } &\mid u \mid > k. \\
\end{array} \right.
\end{equation}

Note that, in general, choosing a per-item turning parameter $k_i$ implies that each item has a different loss function, $\psi_i$. This situation was not directly addressed by Theorem 1. However, Assumptions A1-A4 continue to hold when using the bi-square function with per-item tuning parameters. In particular, $\psi'(0) = 1$ is a constant that does not depend on the choice $k$, so that A3 is satisfied. The bisquare function in Equation \e{psi} is used in the simulation studies and empirical example reported below. 

\subsection{Summary} 

In the case of item intercepts, the R-DIF procedure is defined by Equation \e{M} with $Y_i$ given in Equation \e{Y} and $s_i = \tau_i$ given in Equation \e{tau}. The loss function $\psi$ is defined through assumptions A1-A7, with per-item tuning parameter $k_i$ chosen to be the $1-\alpha/2$ quantile of $N(0, \omega_i$), where $\alpha$ is the desired (asymptotic) Type I Error rate for DIF detection and $\omega_i$ is given in Equation \e{omega}. The resulting estimate of $\theta$ will be denoted $\tilde \theta_{RD}$, and setting $\tilde \theta = \tilde \theta_{RD}$ in part (b) of Theorem 1 will be referred to as the R-DIF test. For computational purposes, Tukey's bi-square in Equation \e{psi} will be used.

%-----------------------------------------------------------------------------------------
\section{Robustness}  

The purpose of this section is to characterize the robustness of the R-DIF procedure in terms of its breakdown point. Theorem 2 shows that the finite sample breakdown point (FSBP) of $\tilde \theta_{RD}$ is determined by the choice of a preliminary estimate of $\theta$, say $\theta^{(0)}$. In practice, this translates into choosing the starting value for iterative estimation procedures such as Newton-Raphson or iteratively re-weighted least squares (IRLS). In particular, choosing $\theta^{(0)}$ to be the median of the $Y_i$ ensures that $\tilde \theta_{RD}$ has the maximum attainable FSBP of any non-trivial location estimator, which is 1/2 \citep{Huber1964}. The section also discusses the implications of breakdown for DIF analysis more generally. 

It is important to emphasize that analytical results on breakdown are quite weak -- they merely describe the minimum proportion of corrupted data points (i.e., items with DIF) that can lead an estimator to take on ``arbitrarily large aberrant values'' \cite[][p.279]{Huber2009}. In practice, the usefulness of any statistic will come into question before it becomes unbounded. Thus, while the concept of breakdown provides a widely used analytical tool for describing robustness, it is helpful for theoretical results to be complemented by numerical examples that characterize how a procedure performs on the way to breakdown. The first simulation study reported in this paper plays this role.   

\subsection{FSBP of the R-DIF procedure}

The definition of FSBP is briefly reviewed before presenting Theorem 2. Let $\bs{Y} = (Y_1, \dots, Y_m)$ denote a sample of size $m$ and $S(\bs{Y}) \in \mathbb{R}$ denote a statistic of interest. In this present context the index $i$ is over items (not respondents), which is why consideration of the FSBP, rather than its asymptotic analogs, is especially relevant. Consider the situation where $\bs{Y}$ is corrupted by replacing $n \leq m$ observations with arbitrary values. The corrupted data can be written as $Y'_i = Y_i + \Delta_i$ with $\Delta_i \in \mathbb{R}$, and $\Delta_i = 0$ for $m - n$ values of $i$. Then $\epsilon = n / m$ is the fraction of corrupted values in $\bs{Y}' = (Y'_1, \dots, Y'_m)$. 

The maximal finite sample ``bias'' in $S$ that can be caused by replacing $\bs{Y} $ with an $\epsilon$-corrupted dataset $\bs{Y}'$ is defined as \cite[see][chap. 11]{Huber2009}:  
\begin{equation}\label{maxbias} 
b(\epsilon, S, \bs{Y}) = \underset{\bs{Y}'}{\sup} \{ \mid S(\bs{Y}) - S(\bs{Y}') \mid  \}
\end{equation}
and the FSBP of $S$ is 
\begin{equation} \label{fsbp}
\epsilon^* = \inf \{ \epsilon \mid b(\epsilon, S, \bs{Y}) = \infty \}.  
\end{equation}
This translates roughly to the smallest proportion of outliers that can cause a statistic of interest to take on arbitrarily large aberrant values. 

In order to derive the FSBP of $\tilde \theta_{RD}$, it is helpful to consider its \emph{one-step} formulation. The intuition behind one-step M-estimation is to start with an initial estimator $\theta^{(0)}$ and update it by applying Newton's rule to Equation \e{M} just once: 
\begin{equation} \label{one-step}
\theta^{(1)} = \theta^{(0)}  -  \frac{\Psi(\theta^{(0)})}{\Psi'(\theta^{(0)})}.
\end{equation}
The one-step estimator appears in the asymptotic theory of M-estimation \citep[see][\S 5.7]{Vaart1998}. In the present context, its utility is to prove the following theorem.

%-----------------------------------------------------------------------------------------
\begin{theorem}
Let $\epsilon^*_r$ denote the FSBP of $\theta^{(r)}$, $r = 0, 1$, as defined by Equations \e{M} and \e{one-step}. Let $\psi(u)$ be defined by assumptions A1-A7, with $u = U_i = (Y_i - \theta) / \tau_i$, $Y_i$ given in Equation \e{Y}, $\tau_i$ given in Equations \e{tau}, and tuning parameters $k_i > 0$. Finally, assume that $\theta^{(0)}$ is not a stationary point of $\Psi(\theta)$. Then $\epsilon^*_1$ =  $\epsilon^*_0$.  
\end{theorem}
%-----------------------------------------------------------------------------------------
 
The proof is given in the Appendix. As mentioned, it depends mainly on treating $\tau_i$ as a known function of $\theta$ (see Equation \e{tau}). It is seen to trivially extend to further iterations, leading to the corollary that $\theta^{(r+1)}$ has the same FSBP as $\theta^{(0)}$ for finite values of $r = 0, 1, 2, \dots$. The assumption that the initial value of $\theta$ is not a stationary point of $\Psi(\theta)$ is restrictive for redescending $\psi$. However, in practice, there are alternative estimation procedures that do not require this assumption (e.g., IRLS). 

Theorem 2 is not directly addressed by past research, although other authors have mentioned the importance of using robust starting values for redescending M-estimators \cite[e.g.,][\S 2.8.1]{Maronna2019}. Huber (\citeyear{Huber1984}) considered redescending M-estimators of location in which the median absolute deviation of the $Y_i$ is used in place of $\tau_i$, showing that $\epsilon^*$ depends not only on the choice of $k$ but also $\bs{Y}$. In particular, he recommended using $k = 6$ in order to ensure $\epsilon^* \approx 1/2$. Li and Zhang (\citeyear{Li1998}) showed that the large sample breakdown of redescending M-estimators can be substantially lower than 1/2. For example, using their results with the value of $k =1.96$ (i.e., the .975 percentile of the standard normal), the breakdown point of the bisquare is expected to be less than 1/3. These considerations suggest that tuning redescending M-estimators to aggressively flag outliers will, paradoxically, come at the cost of robustness. The intuition here is that redescending functions with small values of $k$ can omit substantial portions of the data and therefore lead to local ``bad'' solutions. Yohai (\citeyear{Yohai1987}) showed that the FSBP of redescending M-estimators can be as high as 1/2 when $\tau = \tau_i$ is instead chosen as the solution to a preliminary M-estimation problem. Theorem 2 is in a similar vein, although in this case the result depends on treating $\tau_i$ as a known function of $\theta$ (i.e., Equation \e{tau}), which is a peculiar aspect of the IRT scaling problem. 

%The upshot of Theorem 2 is that the breakdown of R-DIF is determined by the choice of starting value, which is discussed in more detail in the following section. However, we have yet to consider how the breakdown of $\tilde \theta_{RD}$ impacts the R-DIF test statistic $T_i$  defined in part (b) of Theorem 1. This can be addressed through the following inequality:
%%
%\[ T_i >  \frac{Y_i - \tilde \theta}{\sqrt{\tau_i(\tilde \theta)}}. \]  
%%
%Algebraic manipulation of Equation \e{tau} shows that $|\theta| / \sqrt{\tau_i(\theta)} \rightarrow \infty$ as $|\theta| \rightarrow \infty$. On the other hand, for the ``good'' data points in an $\epsilon$-corrupted dataset (i.e., for $\Delta_i = 0$) we have $Y'_i = Y_i$ for some fixed value $Y_i$, so that $Y_i / \sqrt{\tau_i(\theta)} \rightarrow 0$ as $|\theta| \rightarrow \infty$. Consequently, breakdown of $\tilde \theta$ implies that $|T_i| \rightarrow \infty$ for the $m(1 - \epsilon^*)$ uncorrupted values of $Y_i$, or a ``finite sample Type I Error rate'' of 1. 

%-----------------------------------------------------------------------------------------
\subsection{Breakdown, IRT-based scaling, and DIF}

Before moving on, let us briefly consider some more general implications of the concept of breakdown in IRT-based scaling and DIF analysis. In particular, the concept of ``worst-case" DIF is introduced and used to motivate an informal argument against the existence of a (non-trivial) DIF detection procedure with FSBP $ > 1/2$. This overall rationale is also used to design the first simulation study reported below. 

Let $\bar Y = \sum Y_i / m$ (i.e., $\psi(u) = u$ and $s_i = 1$) be the unweighted average of the scaling functions $Y_i$. As defined in Equation \e{maxbias}, the finite sample bias is of $\bar Y$ is $| \bar Y - \bar Y' | = \sum_{i=1}^m \Delta_i / m $. Thus, for any fixed values of $\epsilon$ and $\Delta^* = \max \{ \Delta_i\}$, the maximum bias results when all $\Delta_i > 0$ are set equal to $\Delta^*$. Otherwise stated, the worst-case bias in the estimated scaling parameter will result when all items with DIF are biased in the same direction by the same (maximal) amount. The overall logic of this argument can be extended to other functions used for IRT-based scaling, which all involve unweighted sums of over items \cite[][\S 6.3]{Kolen2014}. The term ``worst-case'' DIF will be used to mean that DIF is not only in the same direction but also by the same magnitude. Worst-case DIF is a special case of unbalanced DIF, which occurs when all items with DIF are biased in the same direction, but not necessarily by the same magnitude \citep[e.g.,][]{Sireci2013}.

Consideration of worst-case DIF leads to the following informal argument against the existence of a (non-trivial) DIF detection procedure with FSBP $> 1/2$. If exactly $\epsilon = 1/2$ of the items on a test exhibit worst-case DIF, then there are two equivalent ways to identify the IRT model in Equation \e{IRT}. To see this, let $\mathcal I = \{i \mid \Delta_i = 0\}$ denote the items without DIF and let $\mathcal J = \{j \mid \Delta_j = \Delta^*\}$ denote the items with DIF. For notational convenience, let us also assume that the item slopes are equal to one for all items in both groups. Then the ``correct'' parameterization of the IRT model in Equation \e{IRT} is obtained by setting $d_{1i} = d_{0i}$ for $i \in \mathcal I$ and $d_{1j} = d_{0j} + \Delta^*$ for $j \in \mathcal J$. However, transforming the latent trait as $\eta^* = \eta - \Delta^*$ and the item parameters as $d^*_{1i} = d_{1i} - \Delta^*$ results in the same IRT model equations, but now $d^*_{1i} = d_{0i} - \Delta^*$ for $i \in \mathcal I$ and $d^*_{1j} = d_{0j}$ for $j \in \mathcal J$ -- i.e., the items with and without DIF have ``flipped.''

This argument is valid for $0 < \epsilon < 1$. However, when $\epsilon = 1/2$, the problem is especially vexing because we cannot use the number of items with DIF to judge the better parameterization. In the absence of an external criterion that can be used to judge which items have DIF, this informal argument suggests FSBP $= 1/2$ is the best we can hope for in DIF analysis. 

%The purpose of this toy example is to consider the \emph{worst-case} scenario, and, in doing so, provide some assurances about how statistical procedures for assessing DIF will perform in more realistic situations. This logic is also used to design the simulation studies reported below. 

%-----------------------------------------------------------------------------------------
\section{Estimation}

This section addresses computational aspects of R-DIF. These procedures are implemented in the \texttt{robustDIF} package (\href{https://github.com/peterhalpin/robustDIF}{https://github.com/peterhalpin/robustDIF}), 
written in the R language \citep{R}. The package defaults discussed in this section were established through unreported simulation studies, although the defaults can be overridden by the user. 

Estimation of $\tilde \theta_{RD}$ can proceed using known results. In particular, Newton-Raphson IRLS can be easily implemented for M-estimators of location \cite[][\S 6.7]{Huber2009}. Location problems typically proceed by treating $\tau_i = \tau$ as fixed to some initial value (e.g., the median absolute deviation) that is not updated during estimation. For R-DIF, we can instead compute $\tau_i^{(0)} = \tau_i(\theta^{(0)})$ for some initial estimate $\theta^{(0)}$. Alternatively, after each iteration $r = 0, 1, \dots$, the updated values $\tau_i^{(r)}$ can be used while solving for $\theta^{(r+1)}$. In \texttt{robustDIF}, the default estimator is IRLS with $\tau_i^{(r)}$ updated during estimation. 

As indicated by Theorem 2, the choice of starting value is important for ensuring the robustness of the R-DIF procedure. The median of the $Y_i$, denoted $\text{med}(Y)$, is a good choice \citep{Huber1964}. However, there are other good choices as well. The least trimmed squares estimator with 50\% trimming rate, denoted $\text{LTS}_{.5}(Y)$, also has FSBP of 1/2 and is straightforward to compute for location problems \citep{Rousseeuw1987}. Additionally, for any choice of $\psi$ such that $\rho = \int \psi(u)du$ exists, one may consider the related problem of minimizing $R(\theta) = \sum_i \rho(U_i)$ with $U_i = (Y_i - \theta) / \tau_i$. In practice, taking the minimum over the grid $\theta \in \Theta_r = \{\min(Y), \min(Y) + r, \dots, \max(Y) - r, \max(Y) \}$ appears to works quite well for $r \leq .05$. In \texttt{robustDIF}, the default starting value is the median of these three choices: 
\[ \theta^{(0)} = \text{med} \{\text{med}(Y), \text{LTS}_{.5}(Y), \underset{\theta \in \Theta_{.05}}{\text{arg min}} \{R(\theta)\}  \} \] 
The user may alternatively choose any one of these, or input their own numerical starting value. 

As previously described, the R-DIF test can be implemented during the estimation of $\tilde \theta_{RD}$ by an appropriate choice of item-specific tuning parameters $k_i$.  One could alternatively follow up the estimation procedure with a ``stand-alone''  test based on part (b) of Theorem 1. These two approaches will be numerically equivalent when $|\theta_i^{(r)} - \theta_i^{(r+1)}| < \delta$ is sufficiently small. The choice of $\delta = 10^{-5}$ is used as the convergence criterion in \texttt{robustDIF}. The R-DIF test can be implemented by flagging items with DIF during estimation or by computing the Wald test in a follow-up step.  

A final note concerns local solutions, which can arise with redescending M-estimators due to the non-monotonicity of $\psi$. The problem can be diagnosed by plotting the function $R(\theta)$ against $\theta$. When there is a clear global minimum, convergence to that minimum can be ensured by the choice of an appropriate starting value, as addressed above. However, it is less clear how to proceed when the are multiple local minima with roughly the same value of $R(\theta)$. One option is to  ``down-tune`` the R-DIF estimator by choosing $k_i$ based on a lower-than-desired Type I Error rate, which has the effect of smoothing out local solutions \citep{Huber1984}. This can be done when choosing starting values based on $R(\theta)$ and also during computation of $\tilde \theta_{RD}$. In the latter case, one may follow up estimation with a stand-alone R-DIF test conducted at the desired Type I Error rate. Another possibility is to report the multiple solutions and weigh their substantive interpretations. 

\section{Extensions} 

Up to this point, the focus has been on the item intercepts. Extension to the item slopes can be made by using Equations \e{CINEG1} and \e{CINEG3} to write
\begin{equation} \label{Y2}
Z_i(\bs{\nu}) =  a_{1i}/a_{0i} ,
\end{equation}
with $Z_i(\bs{\nu}) = \sigma$ representing the null hypothesis of no DIF on the slope of item $i$. Under this null hypothesis, Equation \e{grad} is replaced by
\begin{align} \label{gradZ} \notag
 \nabla Z_i(\bs{\nu}_i) & = a_{0i}^{-1} \left[ -Z_i(\bs{\nu}), 0 , 1,  0 \right]^T\\
			          & = a_{0i}^{-1}  \left[ -\sigma, 0 , 1,  0 \right]^T. 
\end{align}
which leads to the following expression for the variance of the asymptotic null distribution of $Z_i = Z_i(\hat{\bs{\nu}})$ : 
\begin{equation} \label{varZ} 
\text{var}(Z_i ) = a_{0i}^{-2} \left( \sigma^2 \, \text{var}(\hat a_{0i}) + \text{var}(\hat a_{1i}) \right).
 \end{equation}
Note that, similarly to the case of item intercepts, the null hypothesis allows us to write the variance of the null distribution in terms of the target parameter $\sigma$. From here, setting $Y_i(\bs{\nu}) = Z_i(\bs{\nu})$, $\theta = \sigma$, and $\tau_i(\theta) = \text{var}(Z_i )$ in the forgoing sections shows that the same developments carry through to the case of item slopes. 

If the distribution of $Z_i$ is strongly skewed this can lead to problems estimating its location \cite[][\S5.1]{Huber2009}. In such cases, it can be preferable to instead work with $ \log Z_i(\bs{\nu})$. Then the target scaling parameter becomes $\theta = \log \sigma$ and, under the null hypothesis, $\text{var}(\log Z_i) = \text{var}(Z_i ) / \sigma^{2}$. In general, it is recommended to examine the empirical distribution of $Y_i$ and $Z_i$ to determine whether it may be more suitable to work with their log (or another) transformation. 

In addition to using part (b) of Theorem 1 to test the slopes and intercepts of item $i$ separately, one may test them simultaneously using the quadratic form
\begin{equation} \label{chi1}
Q_i = \left[\begin{array}{cc}  Y_i - \tilde \theta_{RD}     &  Z_i - \tilde \sigma_{RD}   \end{array} \right]
\; \Sigma_{i}^{-1} \; 
\left[\begin{array}{c}  Y_i - \tilde \theta_{RD}     \\   Z_i - \tilde \sigma_{RD}   \end{array} \right],
\end{equation}
where the covariance matrix $\Sigma_i$ is given by 
\begin{equation} \notag
\Sigma_i = \left[\begin{array}{cc}  \text{var}(Y_i - \tilde \theta_{RD}) &       \text{cov}(Y_i - \tilde \theta_{RD}, Z_i - \tilde \sigma_{RD}) \\ 
\text{cov}(Y_i - \tilde \theta_{RD}, Z_i - \tilde \sigma_{RD}) & \text{var}(Z_i - \tilde \sigma_{RD})
\end{array} \right].
\end{equation}
Under the joint null hypothesis that none of the item slopes or intercepts exhibit DIF, the variances are obtained from part (b) of Theorem 1. Following the same steps taken in the Appendix, the ``null covariance'' is shown to be 
\begin{equation} \label{chi2}
\text{cov}(Y_i - \tilde \theta_{RD}, Z_i - \tilde \sigma_{RD}) = \sum_{j = 1}^{m} \tilde w(Y_j) \tilde w(Z_j)\, \text{cov}(Y_j, Z_j) 
\end{equation}
with
\begin{equation} \label{chi3} 
\text{cov}(Y_j, Z_j) = \frac{1}{a_{0j}a_{1j}} \left(\sigma \,  \text{cov}(\hat{a}_{0i}, \hat{b}_{0i})  + \text{cov}(\hat{a}_{1i}, \hat{b}_{1i})  -\theta\, \text{var}(\hat{a}_{1i})   \right), 
\end{equation}
\begin{equation} \label{chi4} 
 \tilde w(U_j) = \left\{\begin{array}{ccc}
	 1 - w(U_j) & \text{ for } &i = j  \\
	  w(U_j) & \text{ for } & i \neq j 
\end{array} \right. ,  
\end{equation}
and
\begin{equation} \label{chi5} 
 w(U_j) = \frac{1 / \text{var}(U_j)}{\sum_{k=1}^{m} 1 / \text{var}(U_k)}
 \end{equation}
Then simultaneously testing for DIF on both the slope and intercept of item $i$ can proceed via a Wald test of $Q_i$.  The availability of Equations \e{chi1} through \e{chi5} implies that this test does not require simultaneous estimation of $\theta_{RD}$ and $\sigma_{RD}$. Thus, the Wald test of $Q_i$ can be conveniently implemented as a follow-up to estimation of the individual IRT scaling parameters, as described in the previous section.  

%-----------------------------------------------------------------------------------------
\section{Numerical Examples} 

This section presents two simulation studies illustrating the R-DIF procedure. The first addresses its breakdown in the presence of worst-case DIF. The second addresses statistical power when only a single item exhibits DIF. The simulation studies are followed by a real data example from cross-cultural human development. The example data are publically available from UNICEF \footnote{\url{https://mics.unicef.org/surveys}}. The samples used in the illustration along with R code for running the simulations and conducting the analyses are available at (\href{https://github.com/peterhalpin/robustDIF}{https://github.com/peterhalpin/robustDIF}). The \texttt{R} package \texttt{mirt} \citep{Chalmers2012} was used for estimation of IRT models, \texttt{difR} was used for the Mantel-Haenzal (MH) test \cite{Magis2010}, and \texttt{GPCMlasso} was used to illustrate a regularization-based approach \cite{Schauberger2020}. A nominal Type I Error rate of .05 was used for all procedures, except the regularization-based approach selects the tuning parameter by minimizing BIC. 

%-----------------------------------------------------------------------------------------
\subsection{Simulation 1: Breakdown} 

Data were generated using the 2PL IRT model in Equation \e{IRT}. The focal factor of the study was the proportion of items with DIF, which ranged from 0 to $1/2$. Items with DIF were simulated by applying a bias of $\Delta = .5$ to the item difficulty parameters (not intercepts) and items with DIF were randomly selected in each replication. The other design factors are summarized in Table \ref{tab:sim1}. Note that the simulation study is intended to reflect the worse-case bias that can be induced for a given maximum effect size $\Delta$. The rationale for this design was discussed in the section of this paper entitled ``Robustness''. 

\begin{table}[ht]
\caption{Summary of Simulation 1 Design.}
\label{tab:sim1}
\begin{tabular}{ll}
  \hline
Design factor & Value \\ 
\hline
Proportion of items with DIF (focal) & 0 to 1/2 \\ 
Number of items & $m = 16$ \\ 
Respondents per group &  $n_0 = n_1 = 500$ \\
Replications per condition & 500 \\
Distribution of latent trait & $\eta_{0j} \sim N(0, 1)$ and $\eta_{1j} \sim N(.5, 1)$ \\ 
Item slopes & $a_{0i} \sim U(.9, 2.5)$ and $a_{1i} = a_{0i}$ \\
Item intercepts (without DIF) &  $b_{0i} \sim U(-1.5, 1.5)$, $b_{1i} = b_{0i}$,  and $d_{gi} = b_{gi}\, a_{gi}$ \\
   \hline 
\end{tabular} \\ \\
\emph{Note}:  DIF on item intercepts was simulated using $b_{1i} + .5$ for randomly selected values of $i$ in each replication. 
\end{table}

In each simulation condition, the performance of R-DIF was compared to two traditional methods of DIF analysis, the MH procedure \cite[MH;][]{Dorans1993} and the likelihood ratio test \cite[LRT][]{Thissen1993}, as well as a more recent method that uses regularization \cite[GPCM-lasso;][]{Schauberger2020}. The MH and GPCM-lasso methods assume uniform DIF, but R-DIF and LRT do not require uniform DIF and were implemented without this assumption. Both MH and LRT were estimated using two-stage purification and refinement. Simulation conditions often resulted in all or no items in the anchor set, in which case purification was not performed. It can be noted that many other choices of anchor items are available \cite[see][]{Kopf2015b}, and the results of this simulation study do not seek to address those other choices. 

\begin{figure}[h]
\centering
{\includegraphics[width=15cm]{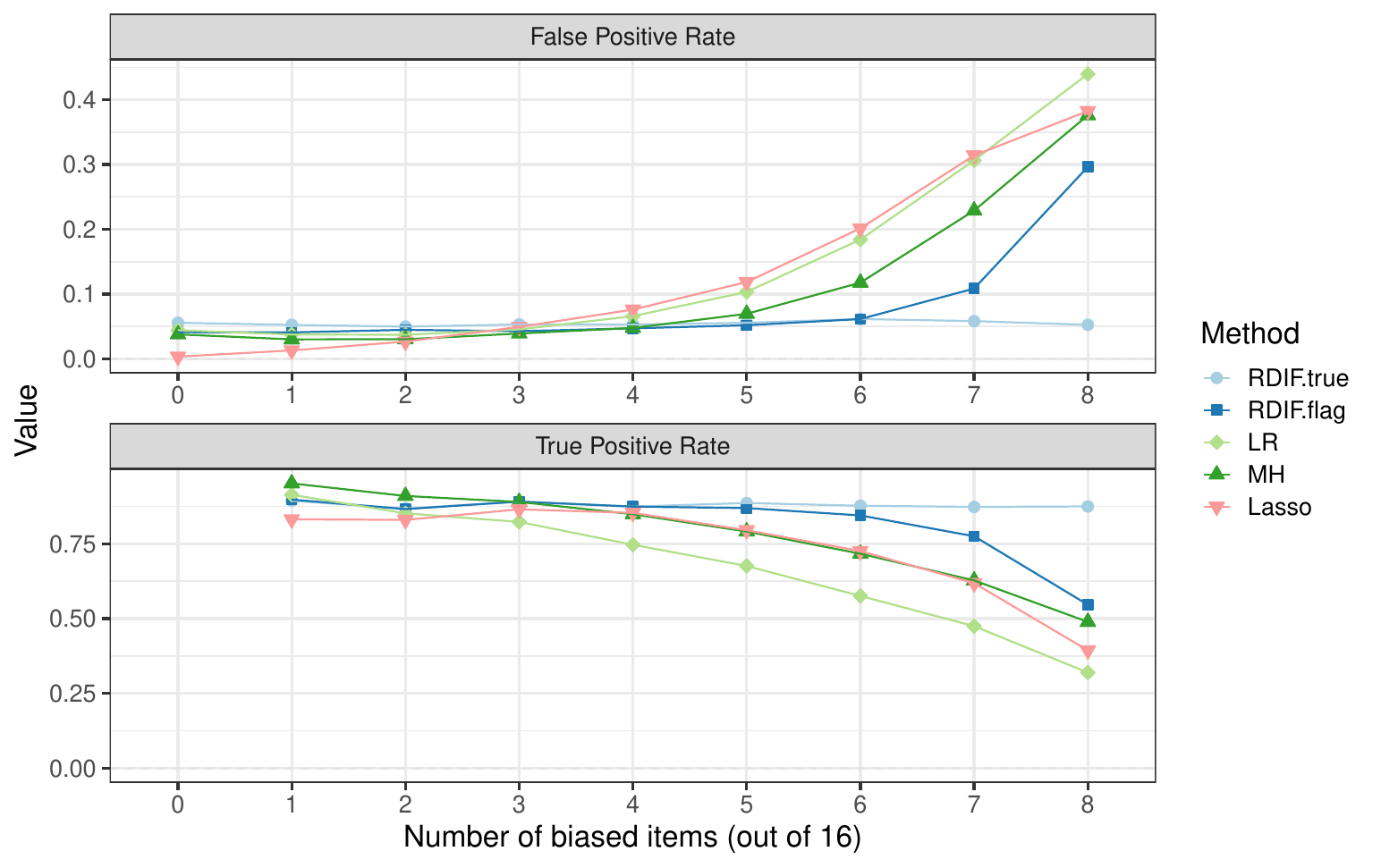}}
\caption{Type I Error rates and statistical power for each of four methods: ``Lasso'' = GPCM lasso; ``MH'' = Mantel-Haenszel; ``LRT'' = LRT; ``RDIF.flag'' = the proposed method.``RDIF.true'' denotes the proposed method computed using the data generating value of $\theta$. }
\label{fig:1}
\end{figure}

The main results are summarized in Figure \ref{fig:1}. The light blue line reports the R-DIF flagging procedure computed using the true scaling parameter. It can be viewed as a check on the correctness of the R-DIF procedure. The dark blue line shows the R-DIF flagging procedure implemented during the estimation of the IRT scaling parameter. It is seen to provide acceptable Type I Error control until 7/16 of the items exhibit worst-case DIF, which is just shy of its theoretical breakdown point of 1/2 biased items. It also maintains its level of statistical power quite well, up to its breakdown point. The stand-alone Wald test for the item intercept (part (b) of Theorem 1) led to identical results and is not reported. The stand-alone Wald test for simultaneously testing the item slope and intercept is also not reported; it was slightly more powerful than the R-DIF flagging procedure, but also led to slightly more Type I Errors.

The MH procedure had better Type I Error control and power than LRT, and the regularization-based approach had false positive rates similar to LRT and power similar to MH. The main observation to be made is that the R-DIF procedure had comparable performance to these alternatives when relatively few items exhibited DIF, and maintained its level of performance much better when larger proportions of items exhibited DIF, up until its theoretical breakdown point of 1/2. When comparing methods, it is also relevant to emphasize that, unlike MH and GPCM-lasso, R-DIF does not require the assumption of uniform DIF -- in this regard, LRT is the only direct comparator. 

\begin{figure}[h]
\centering
{\includegraphics[width=14cm]{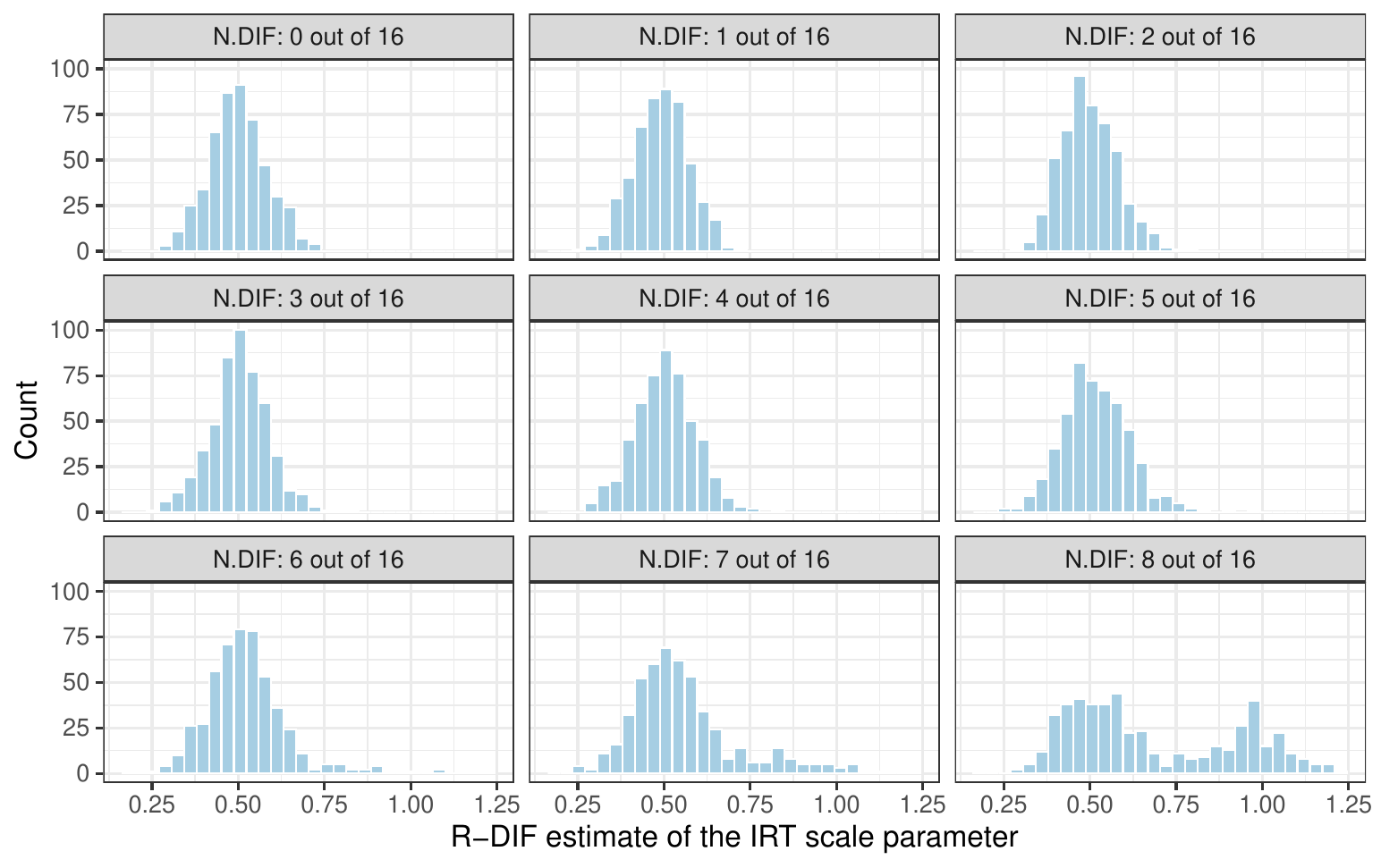}}
\caption{Distribution of $\tilde \theta_{RD}$ in each simulation condition. ``N.DIF'' denotes the number of items with DIF. The data-generating value was 0.5 in each condition. }
\label{fig:2}
\end{figure}

Figure \ref{fig:2} provides another perspective on the breakdown of the R-DIF procedure. The figure shows that the breakdown of the R-DIF test can be explained in terms of the breakdown of the R-DIF estimator of the IRT scaling parameter. %It also provides an empirical illustration of the connection between IRT-based scaling and DIF analysis, as discussed previously. 

%-----------------------------------------------------------------------------------------
\subsection{Simulation 2: Statistical power} 

Data were again generated using the 2PL IRT model in Equation \e{IRT}. This time only a single item exhibited DIF, and the degree of DIF was varied on both the item intercept and item slope. The rationale for limiting consideration to DIF in only a single item is twofold. First, Figure 1 shows that R-DIF maintains its size and power quite well when additional items with the same direction and magnitude of DIF are added. Therefore, consideration of DIF in only a single item provides a reasonable summary of the performance of R-DIF under these more general conditions. Second, focusing on DIF in a single item allows for the statistical power of R-DIF to be fairly benchmarked against traditional methods. In particular, LRT allows for consideration of DIF in both item parameters separately or together, so it is a suitable comparator for R-DIF. But, as shown in Figure 1, LRT does not perform well when additional items exhibit DIF. Thus, limiting DIF to a single item provides a fair way to compare the statistical power of the two methods. 

The simulation design is summarized in Table \ref{tab:sim2}. The focal factors of the study were the sample size per group ($n_0 = n_1 \in \{200, 350, 500 \}$)  and the type of DIF (intercept only, slope only, or both), which were crossed to create nine simulation conditions. In each condition, Type I Error rates and statistical power for R-DIF and LRT were compared, for tests of the intercept only, slope only, and both parameters. For the intercept and slope, R-DIF was implemented by flagging items during estimation. For the two-parameter test, R-DIF was implemented using a follow-up test after estimating each scale parameter separately.  

\begin{table}[ht]
\caption{Summary of Simulation 2 Design.}
\label{tab:sim2}
\begin{tabular}{ll}
  \hline
Design factor & Value \\ 
\hline
Respondents per group (focal)&  $n_0 = n_1 \in \{200, 350, 500 \}$ \\
Type of DIF (focal) & Intercept Only:  $\Delta = .5, \Gamma = 0$ \\ 
	&  Slope Only: $\Delta = 0, \Gamma = 2$ \\ 
	& Intercept and Slope: $\Delta = .35, \Gamma = 1.5$ \\  
Number of items with DIF & 1, randomly selected \\ 
Total number of items & $m = 10$ \\ 
Replications per condition & 500 \\ 
Distribution of latent trait & $\eta_{0} \sim N(0, 1)$, $\eta_{1} \sim N(\mu, \sigma^2)$ with \\ 
	& $\mu \sim U(-.5, .5)$ and $\sigma^2 \sim U(.5, 2)$ \\ 
Item slopes (without DIF) & $a_{0i} \sim U(.9, 2.5)$ and $a_{1i} = a_{0i}$ \\
Item intercepts (without DIF) &  $b_{0i} \sim U(-1.5, 1.5)$, $b_{1i} = b_{0i}$,  and $d_{gi} = b_{gi} a_{gi}$ \\
\hline 
\end{tabular} \\ \\
\emph{Note}:  $\Delta$ denotes additive DIF applied to the item difficulty, $\Gamma$ denotes multiplicative DIF applied to the item intercept.  
\end{table}

Some other aspects of the simulation design warrant mention. The simulation used 10 items, which is the same number as in the real data example reported below. Impact was allowed to vary randomly in each replication, which was intended to make the simulation more realistic. For each type of DIF, the magnitude of DIF was determined so that the non-compensatory DIF index \cite[NCDIFI; e.g.,][Eq. 11]{Raju1995} was approximately equal to 0.1. To compute the NCDIFI, the reference distribution for the latent trait was standard normal and the reference item had a slope of 1 and intercept of 0. DIF on the item intercept was additive and governed by the parameter $\Delta$ applied to the item difficulty, whereas DIF on the item slope was multiplicative and governed by the parameter $\Gamma$. The values of these parameters are given in Table \ref{tab:sim2}. 

\begin{figure}[h]
\centering
{\includegraphics[width=17cm]{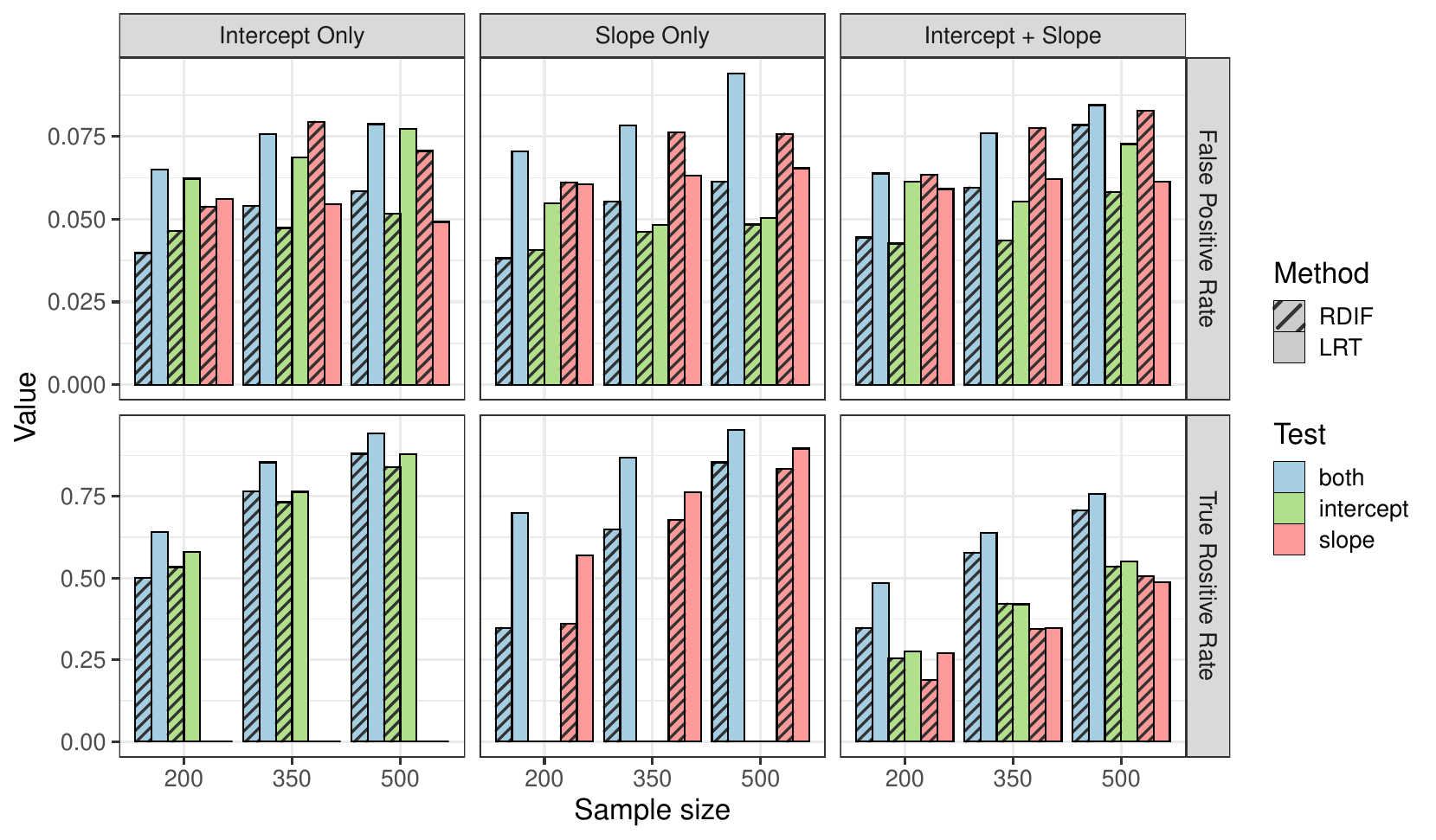}}
\caption{Panels denote the type of DIF (columns) and decision rates (rows). ``Test'' indicates the type of test conducted, with ``both'' denoting the test of both parameters, ``intercept'' denoting a test of the intercepts only, and ``slope'' denoting a test of the slopes only. LRT denotes likelihood ratio tests and RDIF denotes the R-DIF procedure. The nominal Type I Error (false positive) rate of all tests was .05.}
\label{fig:3}
\end{figure}

The results are summarized in Figure \ref{fig:3}. Focussing first on the top row of the figure, it can be seen that all testing procedures maintained the nominal Type I Error rate of .05 reasonably well in all conditions. In particular, note that the R-DIF test for item intercepts was not sensitive to DIF in the item slopes, and vice versa. With a few minor exceptions, the LRT procedure for both parameters had the highest Type I Error rate in all conditions, and was as high as .094 for the largest sample size ($n = 500$ per group) in the Slope Only condition. The overall conclusion is that the Type I Error rate control of R-DIF was comparable to that of LRT.

Turning next to the bottom row of Figure \ref{fig:3}, it can be seen that the R-DIF tests were less powerful than the corresponding LRT test, with only a few minor exceptions. The power differential between R-DIF and LRT was most pronounced when there was DIF on the item slope only (middle panel). In particular, the R-DIF procedure cannot be recommended to test DIF of item slopes with sample sizes less than $n = 350$ per group. In each condition, the power differential between R-DIF and LRT decreased with sample size, suggesting that the differential will become negligible with larger samples.  

%-----------------------------------------------------------------------------------------
\subsection{Empirical example: Assessing human development across countries} 

This section illustrates the use of R-DIF with data from the UNICEF's Early Childhood Development Index (ECDI)\footnote{\href{https://data.unicef.org/resources/early-childhood-development-index-2030-ecdi2030/}{https://data.unicef.org/resources/early-childhood-development-index-2030-ecdi2030/}}. 
The ECDI is a caregiver-reported household survey intended to provide internationally comparable data about the percentage of children aged 24-59 months who are developmentally on track in health, learning, and psychosocial well-being, by sex. Data were collected via household surveys in Fiji and Vietnam. In both surveys, sample frames consisting of a list of households with children between 24 and 59 months were used to design representative samples using probabilistic sampling. The illustration focuses on $m = 10$ ECDI items on the learning domain, which are summarized in the second column of Table \ref{tab:ecdi}, and children aged 48-59 months ($n_0 = 412 $and $n_1 = 978$ in Fiji and Vietnam, respectively). IRT models were estimated using probability-based sampling weights. 

\begin{figure}[h]	
\centering
{\includegraphics[width=12cm]{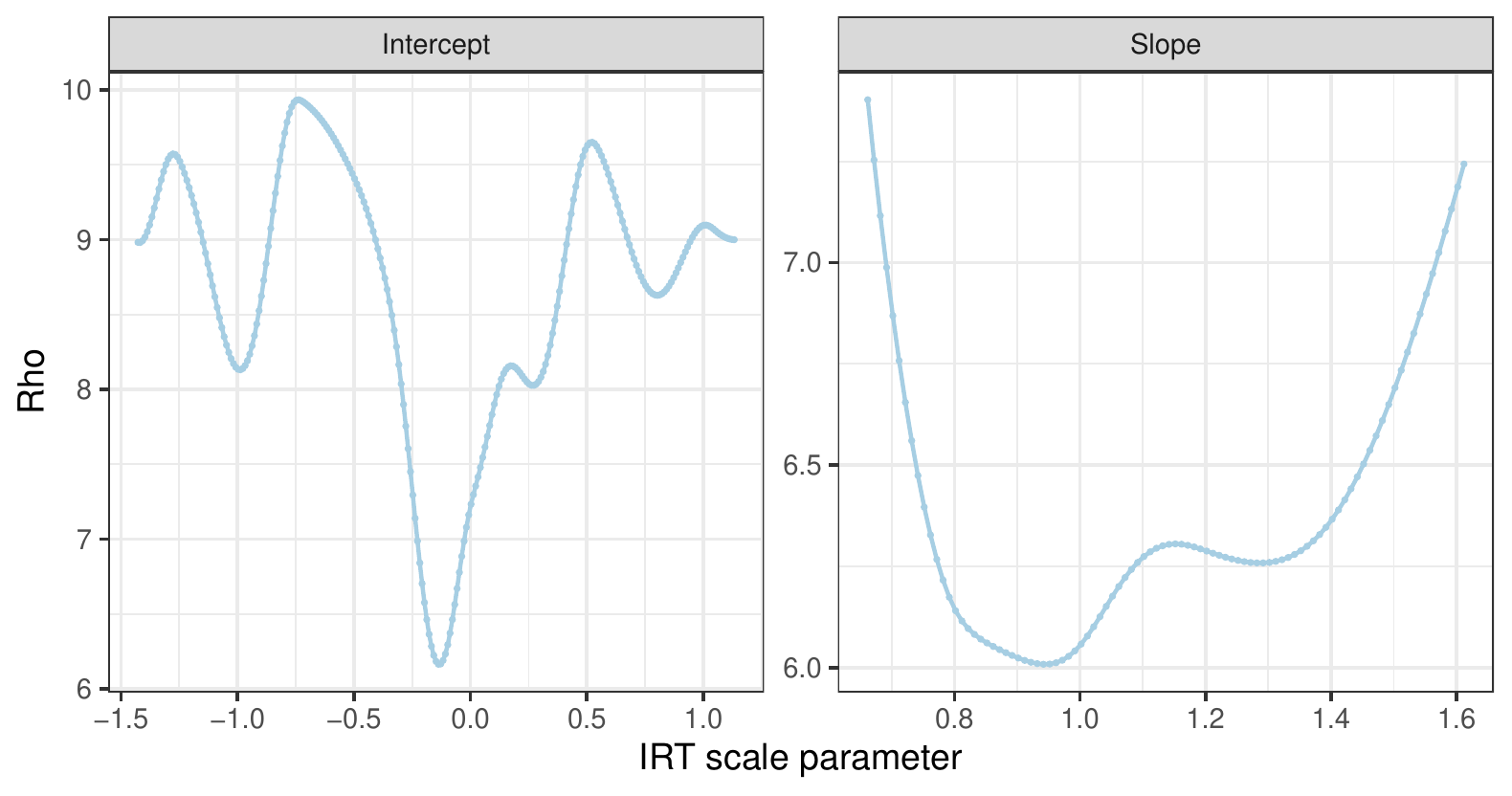}}
\caption{Plots of the $R(\theta)$ minimized by the R-DIF estimator. }
\label{fig:4}
\end{figure}

Figure \ref{fig:4} plots the function $R(\theta)$. As noted in the section of this paper entitled ``Estimation'', $R(\theta)$ is minimized by the R-DIF estimator of $\theta$. The presence of multiple local minima with approximately the same value would indicate potential problems when estimating the IRT scaling parameters and interpreting which items exhibit DIF. However, the figure shows that $R(\theta)$ had clear global minima for both scaling parameters, and the R-DIF procedure converged to these global values. 

Table \ref{tab:ecdi} reports the three types of test statistics available from the R-DIF procedure. Using a Type I Error rate of .05, it was found that six items exhibited DIF on the item intercepts and two items exhibited DIF on the item slopes. The R-DIF test of both parameters led to the conclusion that a total of three items did not exhibit DIF on either parameter. Although the proportion of items with DIF on the intercepts exceeded the theoretical breakdown point of 1/2, DIF was not consistently in the same direction. Combined with the clear global minimum in Figure 4, this suggests that breakdown of the R-DIF procedure was not a concern in the present analysis. 

For comparison, LRT using two-step purification and refinement with a Type I Error rate of .05 led to the conclusion that all items except 4 and 7 exhibited DIF on their intercepts, all items except 4, 6, and 8 exhibited DIF on their slopes, and the test of both parameters identified all items as exhibiting DIF.  The data and code for these analyses are provided at \href{https://github.com/peterhalpin/robustDIF}{https://github.com/peterhalpin/robustDIF}.

\begin{table}[ht]
\caption{R-DIF tests of the ECDI learning items}
\label{tab:ecdi}
\begin{tabular}{llrrrrrr}
\hline
Item & Description & \multicolumn{2}{c}{Intercept} & \multicolumn{2}{c}{Slope} & \multicolumn{2}{c}{Both} \\
\cline{3-4}\cline{5-6}\cline{7-8}
& & $z$ & $p$ value  & $z$ & $p$ value  & $\chi^2$ &  $p$ value \\ 
\hline
1 & Says 10 or more words 		  & 2.55 & \bf{0.01} & 1.60 & 0.11 & 9.89 & \bf{0.01} \\ 
2 & Says sentences of 5 or more words 		  & 3.30 & \bf{0.00} & 2.74 & \bf{0.01} & 11.13 & \bf{0.00} \\ 
3 & Uses correctly “I, you, she, he”			       & 0.24 & 0.81 & -0.56 & 0.57 & 2.13 & 0.34 \\ 
4 & Names an object consistently        & 0.27 & 0.79 & 0.14 & 0.89 & 0.09 & 0.95 \\ 
5 & Recognizes 5 letters of alphabet 			  & -8.18 & \bf{0.00} & 1.85 & 0.06 & 99.14 & \bf{0.00} \\ 
6 & Writes his/her name 		  & -4.97 & \bf{0.00} & 1.29 & 0.20 & 46.59 & \bf{0.00} \\ 
7 & Recognizes all numbers 1 to 5		       & -0.26 & 0.79 & 3.15 & \bf{0.00} & 18.32 & \bf{0.00} \\ 
8 & Gives correct amount (3)   & 0.14 & 0.89 & -0.07 & 0.95 & 0.08 & 0.96 \\ 
9 & Counts to 10  & 3.55 & \bf{0.00} & -1.14 & 0.25 & 20.47 & \bf{0.00} \\ 
10 & Does activities without giving up       & -5.81 & \bf{0.00} & 0.75 & 0.45 & 49.20 & \bf{0.00} \\ 

%1 & Says 10 or more words 			       & -1.58 & 0.11 & 0.39 & 0.69 & 2.70 & 0.26 \\ 
%2 & Says sentences of 5 or more words     & -3.33 & \bf{0.00} & 2.71 & \bf{0.01} & 14.16 & \bf{0.00} \\ 
%3 & Counts to 10 				      	       & 2.87 & \bf{0.00} & -0.20 & 0.84 & 13.94 & \bf{0.00} \\ 
%4 & Recognizes 5 letters of alphabet              & -3.98    & \bf{0.00} & 2.78 & \bf{0.01} & 27.33 & \bf{0.00} \\ 
%5 & Writes his/her name 				      & -1.80 & \bf{0.07} & -2.95 & \bf{0.00} & 42.07 & \bf{0.00} \\ 
%6 & Names an object consistently 		      & 0.95 & 0.34 & 2.78 & \bf{0.01} & 10.25 & \bf{0.01} \\ 
%7 & Says 10 or more words 		              & -3.46 & \bf{0.00} & -2.48 & \bf{0.01} & 11.99 & \bf{0.00} \\ 
%8 & Says sentences of 3 or more words         & 0.18 & 0.85 & 0.20 & 0.84 & 0.04 & 0.98 \\ 
%9 & Says sentences of 5 or more words         & -0.23 & 0.81 & -0.34 & 0.74 & 0.18 & 0.91 \\ 
%10 & Uses correctly “I, you, she, he”             & 3.98   & \bf{0.00} & 3.75 & \bf{0.00} & 22.01 & \bf{0.00} \\ 
\hline 
\end{tabular} \\ \\
\emph{Note}: ``$z$''  denotes the z-test of individual parameters.``$\chi^2$'' denotes the chi-square test of both parameters together and has 2 degrees of freedom. 
The $p$ values are rounded to 2 decimal places and values less than .05 are bolded.  The $p$ values were not adjusted for multiple comparisons. 
\end{table}

%-----------------------------------------------------------------------------------------
\section{Discussion} 

This paper has introduced a method for DIF analysis that is intended for use when a' priori knowledge about anchor items is not available and when many items on an assessment may exhibit DIF. The overall idea is to approach DIF as a problem of outlier detection in IRT-based scaling with the CINEG design. This approach is congenial to M-estimation of a location parameter, with new results providing the asymptotic distribution of the IRT scaling parameters, as well as an asymptotic test of DIF, under the joint null hypothesis that none of the items exhibit DIF. These results were used to develop a highly robust redescending M-estimator that simultaneously provides an estimate of IRT scale parameters and an asymptotic test of DIF, which was referred to as the R-DIF procedure. 

Using the joint null hypothesis that none of the items exhibit DIF to derive asymptotic results about R-DIF may, at first glance, seem to invite the same criticism of logical circularity that has been raised against traditional methods. However, to make use of the joint null hypothesis, the R-DIF procedure requires only that a suitable estimate of the IRT scaling parameters is available. Theoretical results showed that the bias of the R-DIF estimate of the IRT scaling parameters remains bounded so long as fewer than 1/2 of the items on assessment exhibit ``worst-case'' DIF (i.e., biased in the same direction and by the same magnitude). 

The robustness of R-DIF was also illustrated by data simulations, which showed that R-DIF maintains acceptable Type I Error control and statistical power so long as fewer than 1/2 of the item on an assessment exhibit worst-case DIF. While the performance of the comparison methods deteriorated incrementally as more items with DIF were added, R-DIF maintained its initial size and power until approaching its theoretical breakdown point of 1/2. A second simulation study showed that the robustness of R-DIF comes at a cost of reduced statistical power compared to the likelihood ratio test when only a single item exhibits DIF. Thus, R-DIF is most suitable with larger sample sizes ($n \geq 350$ per group).  An empirical example from cross-cultural human development illustrated the use of R-DIF in a context where many assessment items exhibited DIF, and led to substantively different conclusions about DIF compared to the likelihood ratio test. 

This paper focussed on the 2PL model in two independent groups. However, some features of the R-DIF procedure make it suitable for extension. First, the main results presented in this paper trivially extend to other unidimensional psychometric models that (a) can be parameterized in slope-intercept form and (b) have item parameter estimates whose asymptotic distribution is known. This includes, for example, the unidimensional linear factor model and the graded response model. Extensions to wider classes of models (e.g., multidimensional) are less obvious. Second, R-DIF can be implemented using separate calibrations of the focal model in the target populations. Thus it is scalable to situations with many groups, which is especially relevant in cross-cultural settings. The results presented in this paper can be directly applied to pairwise comparisons among multiple groups, although it would be preferable to consider alternative approaches (e.g., sum-to-zero contrasts among groups). A third line of future research is longitudinal settings (i.e., dependent groups). While the asymptotic variances of the IRT scaling parameters given in Equation \e{tau} and \e{varZ} used the assumption that the groups were independent, the main results are agnostic to the specific structure of the asymptotic covariance matrix of the item parameter estimates. 

There are some deeper limitations of the R-DIF procedure that could also be addressed in future research. Most obviously, the methodology relies on asymptotics, which may not always be appropriate. In principle, this limitation can be overcome via bootstrapping, although this complicates the path to analytic results. Second, the concept of an ``effect size'' \cite[e.g.,][]{Sireci2013, Wainer1993}, residual \citep[e.g.,][]{Karabatsos2000, Haberman2009}, or related notions of item misfit  \cite[e.g.,][]{Rost1994, Yamamoto2013} were not addressed in this paper. The focus of the R-DIF procedure is to flag items with DIF, but this leaves open the question of how to quantify the degree of DIF and its consequences for decisions to be made based on test data \cite[e.g.,][]{Chalmers2016, Gonzalez2021}. Developing effect sizes for R-DIF remains an important avenue of future research.  Another limitation concerns the notion of a breakdown point, which provides only a crude characterization of the robustness of an estimator under unspecified types of data contamination. Moving forward, it may be useful to develop more specialized concepts of breakdown that reflect theoretically motivated configurations of DIF and which quantify the consequences of DIF in terms of (finite) degrees of item misfit.  
 
In conclusion, this paper has shown that reframing DIF as a problem in robust scaling can provide a satisfactory resolution to long-standing methodological issues concerning the circular nature of DIF. Consequently, the proposed methodology is especially suited to research settings in which many items may exhibit DIF and anchor items cannot be reliably identified ahead of time. 

%While the present paper focussed on the 2PL IRT model in two independent groups, fruitful lines of future research include extensions to a wider class of psychometric models and research designs, the development of effect sizes to accompany the R-DIF procedure, DIF-specific notions of breakdown, and theoretical results on non-asymptotic inference. 

%-----------------------------------------------------------------------------------------
\section{Appendix} 

\subsection{Proof of Theorem 1}

The proof is obtained via the Delta method \cite[e.g.,][Chap. 3]{Vaart1998}, which requires only assumptions A1 and A4. Assumptions A2 and A3 are used to obtain the distributions of $\tilde \theta$ and $Y_i - \tilde \theta$ under the joint null hypothesis that none of the item intercepts exhibit DIF. In Equation \e{w1} it is seen that Assumption A4 is implied by A2 and A3, so that it is not required to obtain the null distributions (but is required for the non-null distributions). 

The results are organized as follows. First the general (i.e., non-null) asymptotic distribution of $\tilde \theta$ is derived. Then its null distribution is obtained for any choice of $s_i > 0$ in Equation \e{M2}. Finally, the null distribution for $s_i = \text{var}(Y_i)$ is provided. This is followed by an abbreviated version of these same steps for $Y_i - \tilde \theta$. 

For any transformation of the MLEs of the item parameters $g= g(\hat{\bs{\nu}})$ satisfying assumptions A1 and A4, the general form of the result is 
\begin{equation} \label{deltag1}
 \sqrt n \, (g - g(\bs{\nu})) \overset{d}{\rightarrow} N(0, \text{var}(g))
\end{equation} 
where $n = n_0 + n_1$, $n_1 / n_0 = c$ for $c \in (0, \infty)$, and  
\begin{equation} \label{deltag2}
 \text{var}(g) = \nabla g(\bs{\nu})^T \, \text{cov}(\hat{\bs{\nu}}) \, \nabla g(\bs{\nu}). 
\end{equation}
 % 
%
%\begin{equation} \label{gradu}
%\nabla U_i(\bs{\nu})  = \tau_i^{-1} \left(\nabla Y_i(\bs{\nu})  -\nabla \theta(\bs{\nu}) \right) 
%\end{equation}
%
For  $g= \tilde \theta = \theta(\hat{\bs{\nu}})$, the gradient can be obtained by applying the implicit function theorem to Equation \e{M2} (which also requires A4): 
\[ 
\nabla \theta(\bs{\nu}) = - \frac{\partial \Psi}{\partial \bs{\nu}} \left[  \frac{\partial \Psi}{\partial \theta}\right]^{-1}. 
\] 
The required partial derivatives are:
\begin{align} \notag
\frac{\partial \Psi} {\partial \bs{\nu}} 
& = \sum_{i=1}^{m} \psi' \left(U_i(\bs{\nu})\right) \times {\nabla Y_i(\bs{\nu})} / {s_i}
\quad \quad \text{and} \quad\quad 
\frac{\partial \Psi} {\partial \theta}  = - \sum_{i=1}^{m} \psi' \left(U_i(\bs{\nu})\right) / {s_i}.
\end{align}
Substituting these results into Equation \e{deltag2} gives 
\begin{equation}\label{theta1}
\text{var} (\tilde \theta)  = \sum_{i = 1}^{m} w^2_i \, \nabla Y_i(\bs{\nu})^T \, \text{cov}(\hat{\bs{\nu}}) \, \nabla Y_i(\bs{\nu}) = \sum_{i = 1}^{m} w^2_i \, \text{var}(Y_i)
\end{equation}
with weights 
\begin{equation} \label{w1} 
 w_i =  \frac{\psi'(U_i(\bs{\nu}))/s_i} {\sum_{j=1}^{m} \psi'(U_j(\bs{\nu}))/s_j}. 
\end{equation}
Equations \e{deltag1}, \e{theta1} and \e{w1} provide the asymptotic distribution of $\tilde \theta$ for a relatively general specification of $\psi$ (only Assumptions A1 and A4 have been used so far). 

Next, we obtain the distribution of $\tilde \theta$ under the joint null hypothesis that $Y_i(\bs{\nu}) = \theta_0$ for $i = 1, \dots, m$. First it is shown that $\theta(\bs{\nu}) = \theta_0$. Substituting into Equation \e{M2} yields
\begin{equation} 
\Psi(\theta_0, \theta) = \sum_{i=1}^{m} \psi \left(\frac{\theta_0 - \theta}{s_i} \right) = 0. 
\end{equation} 
By assumption A2, $\theta = \theta_0$ is seen to be the unique solution of $\Psi(\theta_0, \theta)$ in a non-empty neighbourhood of $\theta_0$. Thus, under the joint null hypothesis, $\theta(\bs{\nu}) = \theta_0$ and $U_i(\bs{\nu}) = 0$. 

To obtain $\text{var}(\tilde \theta)$ under the joint null hypothesis, note that $\psi'(U_i(\bs{\nu})) = \psi'(0) = c$ and, by assumption A3, $c \neq 0$. So, Assumption A4 is no longer required, and in place of Equation \e{w1} we have the ``null weights'':
\begin{equation} \label{w2} 
 w_i =  \frac{1/s_i} {\sum_{j=1}^{m} 1/s_j}. 
\end{equation}
Thus, for any choice of $s_i$, the joint null hypothesis implies 
\[ 
\sqrt n \, (\tilde \theta - \theta_0) \overset{d}{\rightarrow} N(0, \text{var}_0(\tilde\theta))
\] 
with
\begin{equation}\label{theta2}
\text{var}_0 (\tilde \theta)  =  \sum_{i = 1}^{m} \left(\frac{1/s_i} {\sum_{j=1}^{m} 1/s_j}\right)^2 \text{var}(Y_i).  
\end{equation}

The next step is to choose $s_i$. As mentioned in the preamble to Theorem 1, the goal is to choose the weights to minimize $\text{var}_0 (\tilde \theta) $. Re-writing Equation \e{theta2} using $v_i = w_i \text{var}(Y_i)$ and applying the weighted power means inequality \cite[e.g.,][Chap. 3]{Cvetkovski2012} gives the following lower bound for $\text{var}_0 (\tilde \theta) $: 

\begin{equation} 
\text{var}_0 (\tilde \theta)  =  \sum_{i = 1}^{m} w_i v_i \geq  \left(\sum_{i = 1}^{m} (w_i / v_i)\right)^{-1}  =  \left(\sum_{i=1}^{m} 1/ \text{var}(Y_i)\right)^{-1} 
\end{equation} 

It can be verified that equality is obtained by setting $s_i = \text{var}(Y_i)$ in Equation \e{theta2} which proves part (a) of the theorem. (Incidentally, this is also the variance of the maximum likelihood estimate of $\theta$, which can also be readily verified).

Turning now to part (b), consider the case where $g = Y_i - \tilde \theta$ and $\tilde \theta$ is estimated as just described. Following the same steps outlined above shows that the asymptotic distribution has variance
\begin{equation}\label{e1}
\text{var} (Y_i - \tilde \theta)  = \sum_{j = 1}^{m}  \tilde w^2_j \, \text{var}(Y_j)
\end{equation}
with
\begin{equation} \label{w3}
 \tilde w_{j} = \left\{\begin{array}{ccc}
	 1 - w_j & \text{ for } &i = j  \\
	  w_j & \text{ for } & i \neq j 
\end{array} \right.
\end{equation}
and $w_j$ given by the ``general'' weights in Equation \e{w1}. The null distribution is 
\begin{equation} \label{e2}
\sqrt n \, (Y_i - \tilde \theta) \overset{d}{\rightarrow} N(0, \text{var}_0(Y_i - \tilde \theta))
\end{equation}
with $\text{var}_0(Y_i - \tilde \theta)$ obtained from Equation \e{e1} by using the null weights in Equation \e{w2}.  Finally, setting $s_i = \text{var}(Y_i)$ in Equation \e{w2} and substiting into Equation \e{e1} yields 
\begin{align}\label{e3} \notag
\text{var}_0 (Y_i - \tilde \theta) & =  \text{var}(Y_i)  - 2 w_i\,  \text{var}(Y_i)   + \sum_{j= 1}^{m} w_j^2 \, \text{var}(Y_j) \\ 
& =  \text{var}(Y_i)  - 2\,  \text{var}_0 (\tilde \theta)  +  \text{var}_0 (\tilde \theta). 
%& =  \text{var}(Y_i) - \text{var}_0 (\tilde \theta). 
\end{align}
Equations \e{e2} and \e{e3} provide part (b) of the theorem.

\subsection{Proof of Theorem 2}
Under the assumptions in the theorem, Equation \e{one-step} becomes
\begin{equation} \label{t2}
\theta^{(1)} = \theta ^{(0)} + 
\frac{\sum_i \psi \left( \frac{Y_i - \theta^{(0)}} {\tau_i^{(0)}} \right)}
{\sum_i \psi' \left( \frac{Y_i - \theta^{(0)}} {\tau_i^{(0)}} \right) / \tau_i^{(0)}} 
\end{equation} 
with $\tau_i^{(0)} = \tau_i(\theta^{(0)})$ as given in Equation \e{tau}. Letting the ratio in Equation \e{t2} be denoted by $R$, the theorem requires showing that $|R|$ is bounded away from infinity whenever $|\theta ^{(0)}|$ is. The numerator of $R$ is bounded away from infinity by assumption A6. The denominator is bounded away from zero for finite $\theta^0$ by the assumption that $\theta^0$ is not a stationary point of $\Psi$. These conditions apply to many M-estimators, but leave open the possibility that $R$ will diverge due to $\tau_i^{(0)} \rightarrow \infty$. In the present case, this consideration is eliminated by Equation \e{tau} which shows that $\tau_i(\theta) = O(\theta^2)$. 

\newpage  

\bibliographystyle{apalike}
\bibliography{/Users/halpin/Dropbox/Academic/Manuscripts/Bibliography/library} 

\begin{thebibliography}{}

\bibitem[Angoff, 1982]{Angoff1982}
Angoff, W. (1982).
\newblock Use of difficulty and discrimination indices for detecting item bias.
\newblock In Berk, R., editor, {\em Handbook of {{Methods}} for {{Detecting
  Test Bias}}}, pages 96--116. {The Johns Hopkins Press}, {Baltimore, MA}.

\bibitem[Angoff, 1993]{Angoff1993}
Angoff, W.~H. (1993).
\newblock Perspectives on differential item functioning methodology.
\newblock In Holland, P.~W. and Wainer, H., editors, {\em Differential {{Item
  Functioning}}}, pages 3--23. {Lawrence Earlbaum Associates}, {Hillsdale, NJ}.

\bibitem[Asparouhov and Muth{\'e}n, 2014]{Asparouhov2014}
Asparouhov, T. and Muth{\'e}n, B. (2014).
\newblock Multiple-group factor analysis alignment.
\newblock {\em Structural Equation Modeling: A Multidisciplinary Journal},
  21(4):495--508.

\bibitem[Bechger and Maris, 2015]{Bechger2015}
Bechger, T.~M. and Maris, G. (2015).
\newblock A statistical test for differential item pair functioning.
\newblock {\em Psychometrika}, 80:317--340.

\bibitem[Belzak and Bauer, 2020]{Belzak2020}
Belzak, W. C.~M. and Bauer, D.~J. (2020).
\newblock Improving the assessment of measurement invariance: {{Using}}
  regularization to select anchor items and identify differential item
  functioning.
\newblock {\em Psychological Methods}, 25:673--690.

\bibitem[Bock and Gibbons, 2021]{Bock2021}
Bock, R.~D. and Gibbons, R.~D. (2021).
\newblock {\em Item {{Response Theory}}}.
\newblock {Wiley}, {Hoboken, NJ}.

\bibitem[Chalmers, 2012]{Chalmers2012}
Chalmers, R.~P. (2012).
\newblock Mirt: {{A}} multidimensional item response theory package for the
  {{R}} environment.
\newblock {\em Journal of Statistical Software}, 48(6):1--29.

\bibitem[Chalmers et~al., 2016]{Chalmers2016}
Chalmers, R.~P., Counsell, A., and Flora, D.~B. (2016).
\newblock It might not make a big {{DIF}}: {{Improved}} differential test
  functioning statistics that account for sampling variability.
\newblock {\em Educational and Psychological Measurement}, 76(1):114--140.

\bibitem[Cvetkovski, 2012]{Cvetkovski2012}
Cvetkovski, Z. (2012).
\newblock {\em Inequalities: {{Theorems}}, {{Techniques}} and {{Selected
  Problems}}}.
\newblock {Springer Science \& Business Media}.

\bibitem[Doebler, 2019]{Doebler2019}
Doebler, A. (2019).
\newblock Looking at {{DIF}} from a new perspective: {{A}} structure-based
  approach acknowledging inherent indefinability.
\newblock {\em Applied Psychological Measurement}, 43(4):303--321.

\bibitem[Dorans and Holland, 1993]{Dorans1993}
Dorans, N.~J. and Holland, P.~W. (1993).
\newblock {{DIF}} detection and description: {{Mantel-Haenszel}} and
  standardization.
\newblock In Holland, P.~W. and Wainer, H., editors, {\em Differential {{Item
  Functioning}}}, pages 35--66. {Lawrence Erlbaum Associates}, {Hillsdale, NJ}.

\bibitem[Gonzalez and Pelham, 2021]{Gonzalez2021}
Gonzalez, O. and Pelham, W.~E. (2021).
\newblock When does differential item functioning matter for screening? {{A}}
  method for empirical evaluation.
\newblock {\em Assessment}, 28(2):446--456.

\bibitem[Haberman, 2009]{Haberman2009}
Haberman, S.~J. (2009).
\newblock Use of generalized residuals to examine goodness of fit of item
  response models.
\newblock {\em ETS Reseach Report RR-09-15}.

\bibitem[He, 2013]{He2013}
He, Y. (2013).
\newblock {\em Robust {{Scale Transformation Methods}} in {{IRT True Score
  Equating}} under {{Common-Item Nonequivalent Groups Design}}}.
\newblock {ProQuest LLC}.

\bibitem[He and Cui, 2020]{He2020}
He, Y. and Cui, Z. (2020).
\newblock Evaluating robust scale transformation methods with multiple outlying
  common items under {{IRT}} true score equating.
\newblock {\em Applied Psychological Measurement}, 44(4):296--310.

\bibitem[He et~al., 2015]{He2015}
He, Y., Cui, Z., and Osterlind, S.~J. (2015).
\newblock New robust scale transformation methods in the presence of outlying
  common items.
\newblock {\em Applied Psychological Measurement}, 39(8):613--626.

\bibitem[Huber, 1964]{Huber1964}
Huber, P.~J. (1964).
\newblock Robust estimation of a location parameter.
\newblock {\em The Annals of Mathematical Statistics}, 35(1):73--101.

\bibitem[Huber, 1984]{Huber1984}
Huber, P.~J. (1984).
\newblock Finite sample breakdown of {{M-}} and {{P-estimators}}.
\newblock {\em Annals of Statistics}, 12:119--126.

\bibitem[Huber and Ronchetti, 2009]{Huber2009}
Huber, P.~J. and Ronchetti, E. (2009).
\newblock {\em Robust Statistics}.
\newblock {Wiley}, {Hoboken, NJ}, 2nd edition.

\bibitem[Karabatsos, 2000]{Karabatsos2000}
Karabatsos, G. (2000).
\newblock A critique of {{Rasch}} residual fit statistics.
\newblock {\em Journal of Applied Measurement}, 1(2):152--176.

\bibitem[Kolen and Brennan, 2014]{Kolen2014}
Kolen, M.~J. and Brennan, R.~L. (2014).
\newblock {\em Test {{Equating}}, {{Scaling}}, and {{Linking}}}.
\newblock {Springer}, {New York, NY}.

\bibitem[Kopf et~al., 2015a]{Kopf2015b}
Kopf, J., Zeileis, A., and Strobl, C. (2015a).
\newblock Anchor selection strategies for {{DIF}} analysis: {{Review}},
  assessment, and {{Nnew}} approaches.
\newblock {\em Educational and Psychological Measurement}, 75(1):22--56.

\bibitem[Kopf et~al., 2015b]{Kopf2015a}
Kopf, J., Zeileis, A., and Strobl, C. (2015b).
\newblock A framework for anchor methods and an iterative forward approach for
  {{DIF}} detection.
\newblock {\em Applied Psychological Measurement}, 39(2):83--103.

\bibitem[Li and Zhang, 1998]{Li1998}
Li, G. and Zhang, J. (1998).
\newblock Breakdown properties of location {{M-estimators}}.
\newblock {\em The Annals of Statistics}, 26(3).

\bibitem[Lord, 1980]{Lord1980}
Lord, F.~M. (1980).
\newblock {\em Applications of {{Item Response Theory}} to {{Practical Testing
  Problems}}}.
\newblock {Routledge}, {New York}.

\bibitem[Magis et~al., 2010]{Magis2010}
Magis, D., B{\'e}land, S., Tuerlinckx, F., and De~Boeck, P. (2010).
\newblock A general framework and an {{R}} package for the detection of
  dichotomous differential item functioning.
\newblock {\em Behavior Research Methods}, 42(3):847--862.

\bibitem[Magis et~al., 2015]{Magis2015}
Magis, D., Tuerlinckx, F., and De~Boeck, P. (2015).
\newblock Detection of differential item functioning using the lasso approach.
\newblock {\em Journal of Educational and Behavioral Statistics},
  40(2):111--135.

\bibitem[Maronna et~al., 2019]{Maronna2019}
Maronna, R.~A., Martin, R.~D., Yohai, V.~J., and {Salibi{\'a}n-Barrera}, M.
  (2019).
\newblock {\em Robust {{Statistics}}: {{Theory}} and {{Methods}} (with {{R}})}.
\newblock {Wiley}, {Hoboken, NJ}, 2nd edition.

\bibitem[Mellenbergh, 1982]{Mellenbergh1982}
Mellenbergh, G.~J. (1982).
\newblock Contingency table models for assessing item bias.
\newblock {\em Journal of Educational Statistics}, 7(2):105--118.

\bibitem[{R Core Team}, 2022]{R}
{R Core Team} (2022).
\newblock R: {{A Language}} and {{Environment}} for {{Statistical Computing}}.

\bibitem[Raju et~al., 1995]{Raju1995}
Raju, N.~S., {van der Linden}, W.~J., and Fleer, P.~F. (1995).
\newblock {{IRT-based}} internal measures of differential functioning of items
  and tests.
\newblock {\em Applied Psychological Measurement}, 19(4):353--368.

\bibitem[Robitzsch and L{\"u}dtke, 2023]{Robitzsch2023}
Robitzsch, A. and L{\"u}dtke, O. (2023).
\newblock Why full, partial, or approximate measurement invariance are not a
  prerequisite for meaningful and valid group comparisons.
\newblock {\em Structural Equation Modeling: A Multidisciplinary Journal},
  30(6):859--870.

\bibitem[Rost and {von Davier}, 1994]{Rost1994}
Rost, J. and {von Davier}, M. (1994).
\newblock A conditional item-fit index for {{Rasch}} models.
\newblock {\em Applied Psychological Measurement}, 18(2):171--182.

\bibitem[Rousseeuw and Leroy, 1987]{Rousseeuw1987}
Rousseeuw, P.~J. and Leroy, A.~M. (1987).
\newblock {\em Robust {{Regression}} and {{Outlier Detection}}}.
\newblock {Wiley}, {New York}.

\bibitem[Schauberger and Mair, 2020]{Schauberger2020}
Schauberger, G. and Mair, P. (2020).
\newblock A regularization approach for the detection of differential item
  functioning in generalized partial credit models.
\newblock {\em Behavior Research Methods}, 52(1):279--294.

\bibitem[Sireci and Rios, 2013]{Sireci2013}
Sireci, S.~G. and Rios, J.~A. (2013).
\newblock Decisions that make a difference in detecting differential item
  functioning.
\newblock {\em Educational Research and Evaluation}, 19(2-3):170--187.

\bibitem[Stenhaug et~al., 2021]{Stenhaug2021}
Stenhaug, B., Frank, M.~C., and Domingue, B. (2021).
\newblock Treading carefully: {{Agnostic}} identification as the first step of
  detecting differential item functioning.
\newblock Preprint, {PsyArXiv}.

\bibitem[Stocking and Lord, 1983]{Stocking1983}
Stocking, M.~L. and Lord, F.~M. (1983).
\newblock Developing a common metric in item response theory.
\newblock {\em Applied Psychological Measurement}, 7(2):201--210.

\bibitem[Strobl et~al., 2021]{Strobl2021}
Strobl, C., Kopf, J., Kohler, L., von Oertzen, T., and Zeileis, A. (2021).
\newblock Anchor point selection: {{Scale}} alignment based on an inequality
  criterion.
\newblock {\em Applied Psychological Measurement}, 45(3):214--230.

\bibitem[Thissen et~al., 1993]{Thissen1993}
Thissen, D., Steinberg, L., and Wainer, H. (1993).
\newblock Detection of differential item functioning using the parameters of
  item response models.
\newblock In Holland, P.~W. and Wainer, H., editors, {\em Differential {{Item
  Functioning}}}, pages 67--113. {Lawrence Erlbaum Associates}, {Hillsdale,
  NJ}.

\bibitem[{van der Linden}, 2016]{vanderLinden2016}
{van der Linden}, W.~J. (2016).
\newblock {\em Handbook of {{Item Response Theory}}, {{Volume One}}}.
\newblock {CRC Press}, {Boca Raton, FL}.

\bibitem[van~der Vaart, 1998]{Vaart1998}
van~der Vaart, A.~W. (1998).
\newblock {\em Asymptotic {{Statistics}}}.
\newblock {Cambridge University Press}, {Cambridge, UK}.

\bibitem[{von Davier} and Bezirhan, 2022]{vonDavier2022}
{von Davier}, M. and Bezirhan, U. (2022).
\newblock A robust method for detecting item misfit in large-scale assessments.
\newblock {\em Educational and Psychological Measurement}, page
  00131644221105819.

\bibitem[Wainer, 1993]{Wainer1993}
Wainer, H. (1993).
\newblock Model-based standardized measurement of an item's differential
  impact.
\newblock In Holland, P.~W. and Wainer, H., editors, {\em Differential {{Item
  Functioning}}}, pages 123--135. {Lawrence Erlbaum Associates}, {Hillsdale,
  NJ}.

\bibitem[Wang et~al., 2022]{Wang2022}
Wang, W., Liu, Y., and Liu, H. (2022).
\newblock Testing differential item functioning without predefined anchor items
  using robust regression.
\newblock {\em Journal of Educational and Behavioral Statistics},
  47(6):666--692.

\bibitem[Yamamoto et~al., 2013]{Yamamoto2013}
Yamamoto, K., Khorramdel, L., and {von Davier}, M. (2013).
\newblock Scaling {{PIAAC}} cognitive data.
\newblock In OECD, editor, {\em Technical {{Report}} of the {{Survey}} of
  {{Adult Skills}} ({{PIAAC}})}, pages 17.1--17.34. {OECD Publishing}, {Paris}.

\bibitem[Yohai, 1987]{Yohai1987}
Yohai, V.~J. (1987).
\newblock High breakdown-point and high efficiency robust estimates for
  regression.
\newblock {\em The Annals of Statistics}, 15(2):642--656.

\bibitem[Yuan et~al., 2021]{Yuan2021}
Yuan, K.-H., Liu, H., and Han, Y. (2021).
\newblock Differential item functioning analysis without a priori information
  on anchor items: {{QQ}} plots and graphical test.
\newblock {\em Psychometrika}, 86:345--377.

\end{thebibliography}
%\printbibliography
\end{document}